\documentclass[fleqn,11pt]{SelfArx} % Document font size and equations flushed left
\def\micron{$\mu$m}

\usepackage{natbib}
\usepackage{placeins}

\definecolor{color1}{RGB}{0,0,90} % Color of the article title and sections
\definecolor{color2}{RGB}{0,20,20} % Color of the boxes behind the abstract and headings

\JournalInfo{Keck SSC Whitepaper, Feb. 2015} % Journal information
\Archive{} % Additional notes (e.g. copyright, DOI, review/research article)

\PaperTitle{Science with KRAKENS} % Article title

\Authors{Benjamin A. Mazin\textsuperscript{1}*, 
George Becker\textsuperscript{2}, 
Kevin France\textsuperscript{3}, 
Wesley Fraser\textsuperscript{4,12}, 
D. Andrew Howell\textsuperscript{5,1}, 
Tucker Jones\textsuperscript{1}, 
Seth Meeker\textsuperscript{1}, 
Kieran O'Brien\textsuperscript{6}, 
Jason X. Prochaska\textsuperscript{7}, 
Brian Siana\textsuperscript{8}, 
Matthew Strader\textsuperscript{1}, 
Paul Szypryt\textsuperscript{1}, 
Shriharsh P. Tendulkar\textsuperscript{9}, 
Tommaso Treu\textsuperscript{10}, 
Gautam Vasisht\textsuperscript{11} 
} % Authors
\affiliation{\textsuperscript{1}\textit{Department of Physics, University of California, Santa Barbara, CA, USA }} % Author affiliation
\affiliation{\textsuperscript{2}\textit{Space Telescope Science Institute, Baltimore, MD, USA}} 
\affiliation{\textsuperscript{3}\textit{Department of Astrophysical and Planetary Sciences, University of Colorado, Boulder, CO, USA}} 
\affiliation{\textsuperscript{4}\textit{Herzberg Institute of Astrophysics, Victoria, BC, Canada}} 
\affiliation{\textsuperscript{5}\textit{Los Cumbres Observatory, Santa Barbara, CA, USA}} 
\affiliation{\textsuperscript{6}\textit{Department of Astrophysics, University of Oxford, Oxford, UK}} 
\affiliation{\textsuperscript{7}\textit{Department of Astronomy and Astrophysics, University of California, Santa Cruz, CA, USA}} 
\affiliation{\textsuperscript{8}\textit{Department of Physics and Astronomy, University of California, Riverside, CA, USA}} 
\affiliation{\textsuperscript{9}\textit{Department of Physics, California Institute of Technology, Pasadena, CA, USA}} 
\affiliation{\textsuperscript{10}\textit{Department of Physics and Astronomy, University of California, Los Angeles, CA, USA}} 
\affiliation{\textsuperscript{11}\textit{NASA Jet Propulsion Lab, Pasadena, CA, USA}} 
\affiliation{\textsuperscript{12}\textit{The Astrophysics Research Centre, Queen’s University Belfast, Belfast, UK}}
\affiliation{*\textbf{Corresponding author}: bmazin@physics.ucsb.edu} % Corresponding author

\newcommand{\keywordname}{Keywords} % Defines the keywords heading name

\Abstract{The Keck science community is entering an era of unprecedented change.  Powerful new instrument like ZTF, JWST, LSST, and the ELTs will catalyze this change, and we must be ready to take full advantage to maintain our position of scientific leadership.  The best way to do this is to continue the UC and Caltech tradition of technical excellence in instrumentation.  In this whitepaper we describe a new instrument called KRAKENS to help meet these challenges.  KRAKENS uses a unique detector technology (MKIDs) to enable groundbreaking science across a wide range of astrophysical research topics.  This document will lay out the detailed expected science return of KRAKENS.}

\begin{document}

\flushbottom % Makes all text pages the same height

\maketitle % Print the title and abstract box

\tableofcontents % Print the contents section

\thispagestyle{empty} % Removes page numbering from the first page

%----------------------------------------------------------------------------------------
%	ARTICLE CONTENTS
%----------------------------------------------------------------------------------------

\section{The KRAKENS Instrument}
% Summarize lack of time domain at Keck + K1DM3
The Keck Observatory is currently the world leader in faint object spectroscopy and laser guide star adaptive optics (LGSAO), and will continue to lead the way with innovative instruments like MOSFIRE and KCWI.  However, time domain astronomy, including both cadence observations and rapidly variable sources, is an increasingly large part of Astronomy.  This trend will only increase with surveys like ZTF and LSST.  This has been recognized by Keck and NSF with the funding of K1DM3, a new deployable tertiary for Keck I.  K1DM3 will enable cadence observing, but the current instrument suite on Keck I is not optimal for rapidly time variables sources.  These sources include transients like gamma-ray burst afterglows, supernovae, tidal disruptions, and EM follow-up of gravitational wave detections, as well as more predictable time variable sources like planet transits, optical pulsars, and compact binaries.

We propose to address the need for time domain capabilities on Keck I with a new instrument based on a technology that has been developed at Caltech, JPL, and UCSB: the UV, optical, and near-IR (UVOIR) Microwave Kinetic Inductance Detector (MKID)~\citep{day03}.  MKIDs have been called the ideal detectors --- like a X-ray microcalorimeter, they count single photons and measure their energy without read noise or dark current, and with nearly perfect cosmic ray rejection.  They have now been demonstrated on sky with 34 observing nights at the Palomar 200" and Lick 120" with the ARCONS instrument~\citep{2013PASP..125.1348M}, and the first papers with MKID data~\citep{2013ApJ...779L..12S,2014MNRAS.tmp..307S} have been published.

%\begin{wrapfigure}{l}{0.5\textwidth}
\begin{figure}[t]
\begin{center}
\includegraphics[width=1.0\columnwidth]{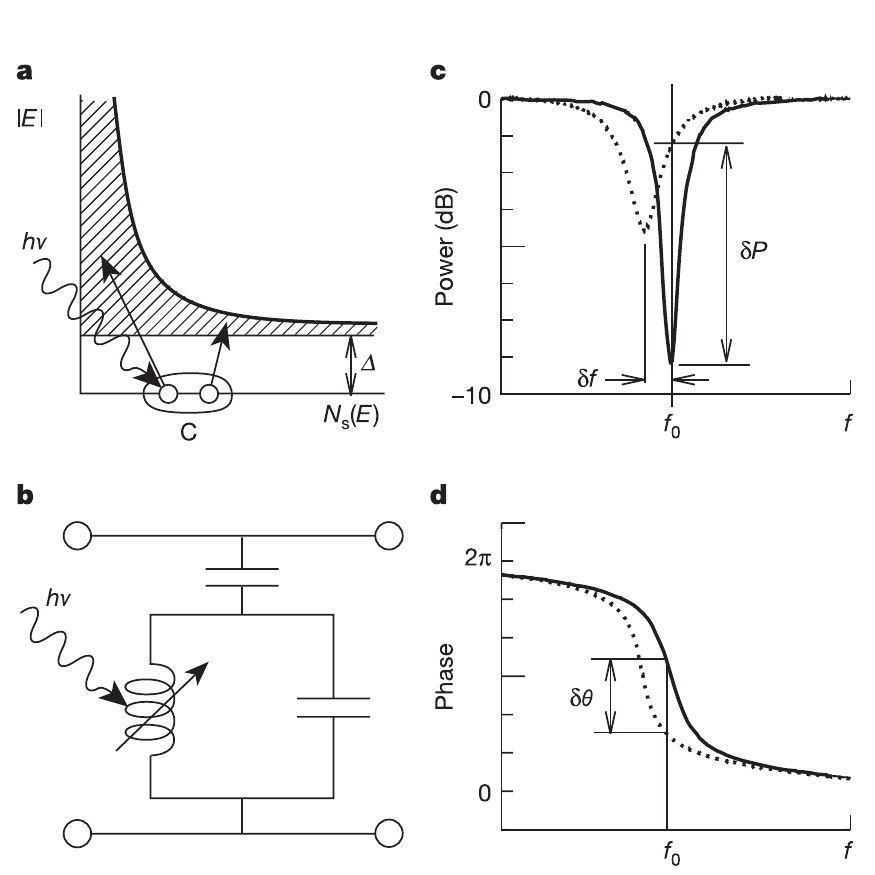}
\end{center}
\vspace{-10pt}
\caption{\footnotesize Left: The basic operation of an MKID, from \cite{day03}. (a) Photons with energy $h\nu$ are absorbed in a superconducting film, producing a number of excitations, called quasiparticles.  (b) To sensitively measure these quasiparticles, the film is placed in a high frequency planar resonant circuit.  The amplitude (c) and phase (d) of a microwave excitation signal sent through the resonator.  The change in the surface impedance of the film following a photon absorption event pushes the resonance to lower frequency and changes its amplitude.  If the detector (resonator) is excited with a constant on-resonance microwave signal, the energy of the absorbed photon can be determined by measuring the degree of phase and amplitude shift.} 
\label{fig:detcartoon}
\vspace{-.2in}
\end{figure}
%\end{wrapfigure}

%\FloatBarrier

MKIDs work on the principle that incident photons change the surface impedance of a superconductor through the kinetic inductance effect.  The kinetic inductance effect occurs because energy can be stored in the supercurrent (the flow of Cooper Pairs) of a superconductor.  Reversing the direction of the supercurrent requires extracting the kinetic energy stored in it, which yields an extra inductance term in addition to the familiar geometric inductance.  The magnitude of the change in surface impedance depends on the number of Cooper Pairs broken by incident photons, and hence is proportional to the amount of energy deposited in the superconductor. This change can be accurately measured by placing a superconducting inductor in a lithographed resonator.  A microwave probe signal is tuned to the resonant frequency of the resonator, and any photons which are absorbed in the inductor will imprint their signature as changes in phase and amplitude of this probe signal.  Since the quality factor $Q$ of the resonators is high and their transmission off resonance is nearly perfect, multiplexing can be accomplished by tuning each pixel to a different resonant frequency with lithography during device fabrication.  A comb of probe signals is sent into the device, and room temperature electronics recover the changes in amplitude and phase~\citep{2012RScI...83d4702M}.  More details on MKIDs can be found in \citep{Mazin:2012kl,2013PASP..125.1348M} and in Figure~\ref{fig:detcartoon}.

%\begin{wrapfigure}{l}{\columnwidth}
\begin{figure}[t]
\begin{center}
%\vspace{-.3in}
\includegraphics[width=\columnwidth]{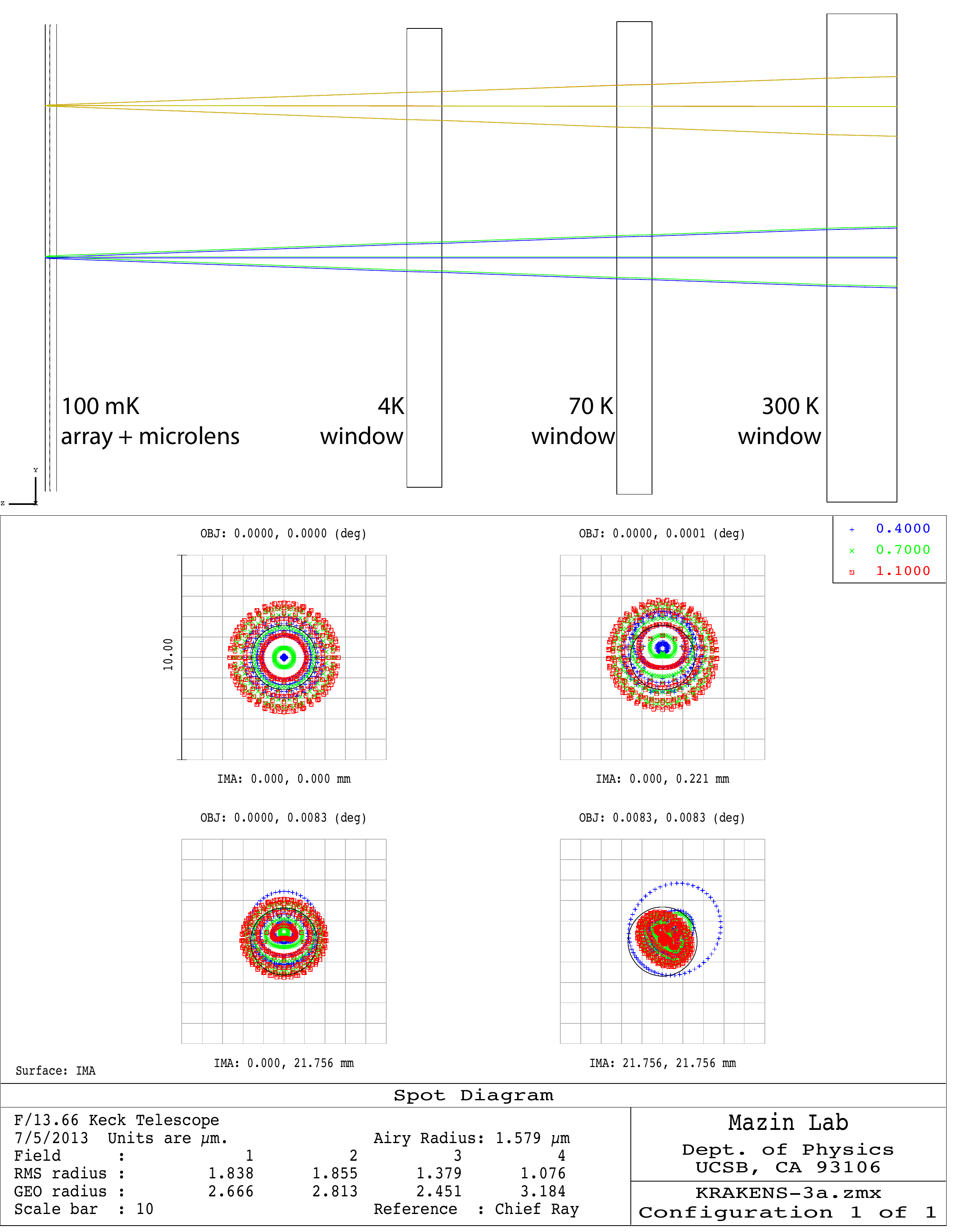}
\end{center}
\vspace{-.1in}
\caption{\footnotesize Top: The simple optical layout of KRAKENS.  Bottom: Spot diagrams of KRAKENS generated with Zemax, including the effect of the microlens array.  KRAKENS imaging performance ill be seeing limited.} 
\label{fig:zemax}
\vspace{-.20in}
%\end{wrapfigure}
\end{figure}
%\FloatBarrier

KRAKENS, the  \textbf{K}eck  \textbf{R}adiometer  \textbf{A}rray using  \textbf{K}ID  \textbf{EN}ergy  \textbf{S}ensors, will build on the knowledge gained from ARCONS to create an instrument of unprecedented capabilities for the Keck I bent Cassegrain port. KRAKENS will be a MKID-based, low resolution, high sensitivity integral field unit (IFU) with 32,400 (180${\times}$180) pixels, a 45${\times}$45 arcsecond field of view, spectral resolution $R=\lambda/\Delta \lambda{\approx}$25 at 0.4~$\mu$m, and an extraordinarily wide 0.32--1.35~$\mu$m bandwidth.  The relatively large FOV will place stable calibration stars on the array for accurate photometry, as well as enabling observations of larger objects like nearby galaxies and galaxy clusters.   KRAKENS will be a facility instrument at Keck, with a full software pipeline inherited from ARCONS, and is expected to cost roughly \$2M. KRAKENS does not require the development of any new technologies.  It is an enlargement and optimization of the technologies already proven in the field with ARCONS.  Two instruments currently funded and under construction, the NSF-funded 10 kpix MKID planet imager DARKNESS and the 20 kpix Japanese-funded planet imager MEC, significantly reducing technical risk.

\textbf{Location:}  The Keck Visiting Instrument Ports, built around a bent Cassegrain f/15 focus, have not been in regular use for the last decade.  The Keck I port was recently refurbished for NIRES, but NIRES is now planned to go on Keck II, leaving a fully commissioned focus open for KRAKENS.  The port includes provisions for both an instrument rotator and a standard Keck offset guider.

\textbf{Optics:} The plate scale with the f/15 secondary is 0.725 mm/arcsec, which is an excellent match to the 0.180 mm pixel size we will use in KRAKENS (ARCONS uses 0.222 mm, and DARKNESS will use 0.150 mm), yielding a plate scale of 0.25 arcsec/pixel with no reimaging optics.  In fact, as shown in Figure~\ref{fig:zemax}, the field curvature of 2.14 meters is large enough over the desired KRAKENS 45" FOV that no field flattener is required.  This means the MKID array can be placed at the direct focus of the telescope with just unpowered vacuum windows and flat IR blocking windows in the beam, allowing extraordinarily high efficiency.  The sole optics in the system, shown in the Figures, consist of an anti-reflection (AR) coated N-BK7 vacuum window, an AR coated N-BK7 77 K IR blocking window, a 4 K 1.35~\micron~short pass filter, and a N-BK7 microlens (92\% optical fill factor) with a backside coating of Indium Tin Oxide to reflect long wavelength IR background.  Given the specs on a very broadband AR coating quoted from Materion of 1.5\% reflection per surface and 90\% transmission from the ITO coating, the KRAKENS optics should have a throughput of over 75\% across the entire 0.35--1.35~\micron~band.  A multilayer bandpass filter will be used to set the red edge of our band, as sky count rate can become problematic, as discussed below.

\textbf{MKID Array:}  The 180${\times}$180 pixel MKID array for KRAKENS will be fabricated at JPL specifically for this project. The array in KRAKENS will be nearly identical to the arrays developed for ARCONS and DARKNESS, with the only difference being a very straightforward upgrade in pixel count.  It is likely (but not required) that continued NASA funding for MKID development will significantly enhance this fabrication effort, improving yield and R from R${\sim}$10 at 0.4~\micron~achieved with ARCONS towards the theoretical limit for a 100 mK operating temperature of R$\sim$100 at 0.4~\micron.  Upgrading KRAKENS with improved MKID arrays will be simple and inexpensive.

%\begin{wrapfigure}{r}{0.2\textwidth}
%\vspace{-.45in}
%\begin{center}
%\includegraphics[width=0.2\columnwidth]{102_model_cryostat.jpg}
%\end{center}
%\vspace{-.2in}
%\caption{The cryostat proposed for KRAKENS.  The total height of the cryostat is roughly 1 m.}
%\label{fig:adr}
%\vspace{-.5in}
%\end{wrapfigure}

%We anticipate the final array will have 222~\micron~pitch pixels.  The active area will be 40${\times}$40 mm, allowing us to fabricate four arrays on a single 150 mm wafer, or 1 array on a 100 mm wafer.  The array will be optimized to give an  R$\sim$20 at 0.4~\micron. This energy resolution is a conservative estimate based on what we are achieving in the lab today.  Adding the new JPL Para-Amp and three years of continuous improvements in the resonator design may significantly increase R towards the theoretical limit for a 100 mK operating temperature of R$\sim$100 at 0.4~\micron.

\textbf{Cryogenics:} Like ARCONS, KRAKENS will be built around an ADR to take it from 4 K to 100 mK.  The decision to use a mechanical pulse tube cooler (like ARCONS) or liquid He cryogen (like DARKNESS) to get to 4 K will be made after more consultation with the Keck engineers, with a strong preference for a mechanical cooler to lower consumable costs.

\textbf{Readout Electronics:}  The readout electronics for KRAKENS will be identical to those for DARKNESS except with three times as many readout boards.  These use a new, Xilinx Virtex-6 based version of the CASPER FPGA platform, the ROACH2.  This allows us to reuse the several man-years of work already spent on firmware development for ARCONS~\citep{2012RScI...83d4702M} and DARKNESS.

\textbf{Fore Modules:}  The default observing mode of KRAKENS is a high throughput, low resolution IFU.  However, in order to increase the flexibility of the camera we intend to implement two deployable Fore Modules.  The first is a polarimeter using a Wedged double Wollaston (WeDoWo)~\citep{Oliva:1997dx} prism or similar.   This will allow us to measure the spectrum and intensity of the polarized flux at 0, 45, 90, and 135 degrees on different part of the array, reducing the FOV by 4 but allowing instantaneous reconstruction of the first three elements of the Stokes vector.

The second Fore Module inserts a grating and potentially a focal plane mask into the beam, reducing the FOV substantially but spreading the light out to achieve a higher R, up to ${\sim}$100 for the entire bandpass and even higher over a narrower wavelength range.  This yields two pieces of information on the energy of every photon --- the absorption position caused by the grating dispersion, and the actual energy recorded by the MKID.  Taking data at multiple grating rotation angles allows us to reduce the background in our slitless grating spectroscopy mode significantly over what can be achieved without an energy resolving array.

\textbf{Adaptive Optics:} Another possible use of KRAKENS is as a very high throughput camera for use behind the LGSAO system.  This would require fore optics to manipulate the plate scale and significant work to smoothly interface with the existing AO system.  The system would work at J band, with the possibility of operation through H band if a cold (4 Kelvin) filter wheel is implemented.  This application will require significant further research to determine the feasibility of the proposed integration.

\begin{figure}
\vspace{-.0in}
\begin{center}
\includegraphics[width=\columnwidth]{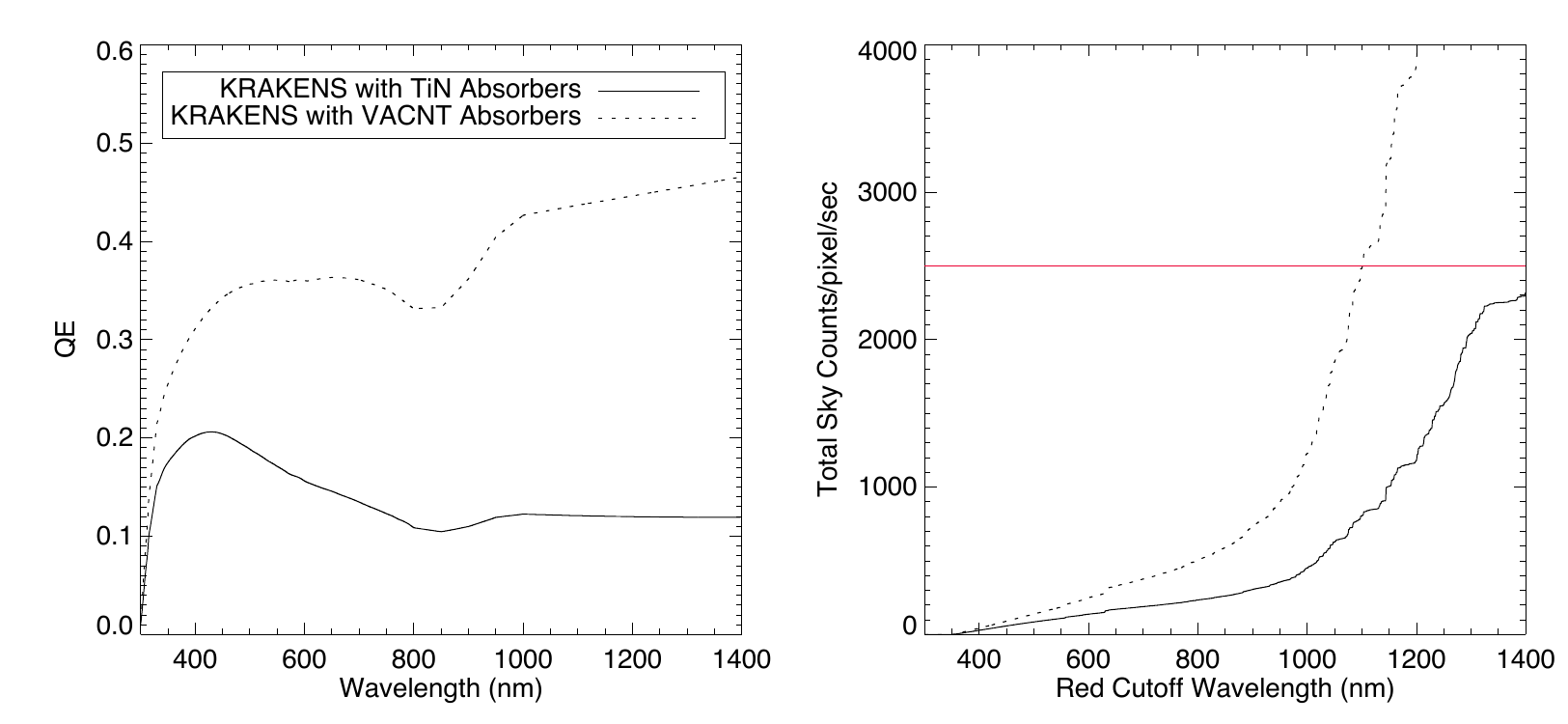}
\end{center}
\vspace{-.2in}
\caption{\footnotesize Left: The predicted total throughput of KRAKENS, including atmosphere, telescope, optics, and the detectors.  Right: The expected sky count rate per pixel as a function of the red cutoff wavelength of our filters.  The effective maximum count rate of 2500 cts/sec/pixel is shown as a red line.  Use of the wavelength dispersive grating fore module will significantly relax this limit.}
\label{fig:qe}
\vspace{-.2in}
%\end{wrapfigure}
\end{figure}

\textbf{Sky Count Rate and Bright Limits:} The predicted total system throughput of KRAKENS, including the atmosphere, telescope, optics, and MKID array is shown in Figure~\ref{fig:qe}.  Two cases are computed, the first with our current MKID QE, and the second assuming successful integration of new absorbers yielding a MKID QE of 80\%.  In the right panel of this figure the expected sky count per pixel for these two cases is computed using a measured high resolution Gemini South sky spectrum normalized to give a V-band magnitude of 21.5 per sq. arcsec.  These plots show that if we surpass the expected sensitivity we will need to set the red end of our band somewhere between 1.0 and 1.25~\micron~to keep the count rate below our maximum, which is set by the physics of the MKID and the readout electronics. 

%\begin{wrapfigure}{r}{\columnwidth}

This count rate limit means that there is a limit on the brightest objects KRAKENS can observe.  With the entire bandpass open, KRAKENS will only be able to observe objects slightly fainter than sky, m$_V{\sim}20$, without defocusing the telescope.  A warm filter wheel with various bandpass and neutral density filters will be employed, allowing for instance 0.32--0.8~\micron~observations of objects to m$_V{\sim}18$.  Even brighter objects will require either neutral density filters, defocusing the telescope, or more likely using the wavelength dispersive grating fore module.  While this restriction does somewhat limit KRAKENS, it is expected that most of the sources brighter than KRAKENS can access would better be served with an instrument on a smaller telescope, like ARCONS.

\begin{figure*}
\vspace{-.0in}
\begin{center}
\includegraphics[width=\textwidth]{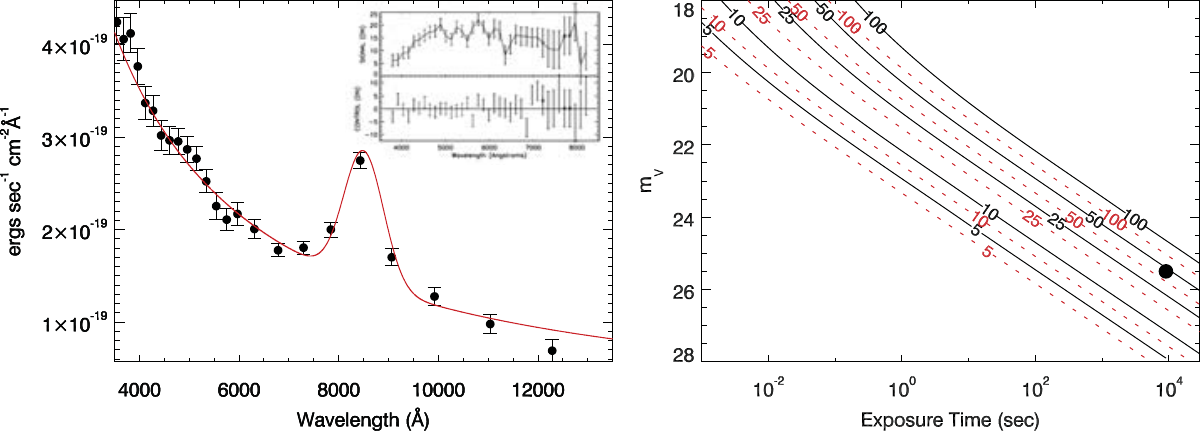}
\end{center}
\vspace{-.2in}
\caption{\footnotesize Left: A 2.5 hour simulated observation (with our lower QE case) of m$_V{\sim}25.4$ optical pulsar Geminga with KRAKENS showing the clear detection of a hypothetical cyclotron emission line at 8500~\AA~with a strength of twice the continuum and a FWHM of 930~\AA. The red line shows the model spectrum. The inset shows the actual spectrum of Geminga taken in a 8 hour observation with LRIS on Keck~\citep{Martin:1998bp}. The KRAKENS spectrum appears superior mainly because of higher throughput and the ability to take advantage of the 0.8" seeing to reduce the sky background rate, wavelength bin size, and better control of systematics such as variations in the intensity and spectrum of the night sky over the 30 minute individual LRIS integrations.  Right: Contours of expected signal-to-noise ratio (SNR) as a function of exposure time and optical magnitude for an object with a Geminga-like power law spectrum $(f_\nu=\nu^{-0.8})$.  The black lines are for KRAKENS observations in 0.8" seeing and a resolution $R=20$ at 0.4~\micron~with our low QE case, while the red dashed are for the high QE case. The black dot represents the 2.5-hr simulation observation of Geminga shown in the left panel.  Note the low integration time performance --- KRAKENS can get a broadband signal to noise ratio of around 100 for a 1 second integration on a $m_V=19$ star.}
\label{fig:perf}
\vspace{-.2in}
\end{figure*}

\textbf{Estimated On-Sky Performance:} The simulated performance of KRAKENS is shown in Figure~\ref{fig:perf}.  KRAKENS can obtain a detection with broadband signal to noise ratio of 25 on a m$_V{=}26$ object in one hour.  KRAKENS is not limited to time variable objects.  Its sensitivity and broad wavelength coverage mean that it could become the instrument of choice for studies of faint and especially low surface brightness objects at Keck.

\textbf{Array Improvements:} The MKID arrays used in KRAKENS take about three man-days to fabricate in the clean room at JPL.  This means the arrays themselves are very inexpensive, with most of the cost residing in the complex room temperature RF electronics.  We will build in a regular, likely yearly, upgrade schedule into the MKID arrays in KRAKENS, allowing significant future upgrades of QE, pixel yield, and spectral resolution as the technology progresses.

\textbf{Risk Management:}  MKIDs are a relatively new technology, and concerns over the data quality are justified.  For instance, detailed studies on stability, sky subtraction, and calibration are ongoing.  However, KRAKENS can't realistically be on the sky until at least 2018B.  By that time we expect to have significantly more experience from our Palomar campaigns with ARCONS, have DARKNESS doing exoplanet imaging at Palomar, MEC doing exoplanet imaging at Subaru with SCExAO, and will likely have flown a MKID-based stratospheric balloon (PICTURE-C).  In addition, groups at Fermilab and Oxford have working UVOIR MKID groups and are seeking funding for instruments.  These issues should be well in hand by the time KRAKENS is commissioned. 

\textbf{Performance Requirements and Goals:}
%\vspace{-.1in}

The performance requirements and goals of KRAKENS are shown in Table~\ref{tab:perf}.  These requirements are the minimum performance required to achieve the scientific goals laid out in this paper.  The goals are higher targets for enhanced performance that currently appear reasonable given the pace of MKID development.

\begin{table*}
\small
\begin{center}
\noindent\begin{tabular}{lccc}
\hline
\hline
\textbf{Name} & \textbf{ARCONS} & \textbf{KRAKENS Req.} & \textbf{KRAKENS Goal}\\
\hline
Plate Scale & 0.45 "/pixel & 0.25 "/pixel & 0.25 "/pixel \\
\hline
Pixel Count & 2024 & 32400 & 57600 \\
\hline
Fraction of Functional Pixels$^a$ & 70\% & 85\% & 95\% \\
\hline
Field of View & 20$\times$20" & 45$\times$45" & 60$\times$60" \\
\hline
Maximum Count Rate & 1000 cts/pixel/sec & 1500 cts/pixel/sec & 2500 cts/pixel/sec \\
\hline
Spectral Resolution$^b$ & R=10 at 0.4 \micron & R=15 at 0.4 \micron & R=25 at 0.4 \micron \\
\hline
Cryostat Optics Throughput & $<$50\% & 65\% & 75\%\\
\hline
Mean MKID Quantum Efficiency$^c$  & 30\% & 30\% & 80\% \\
\hline
Wavelength Coverage$^d$  & 0.4--1.1 \micron & 0.37 -- 1.1 \micron & 0.32 -- 1.35 \micron \\
\hline
Cryogenic Base Temperature Hold Time & 12 hours & 12 hours & 16 hours\\
\hline
\hline
\end{tabular}
\end{center}
\caption{\footnotesize KRAKENS Performance goals and requirements. a: The number of functional pixels is limited by frequency collisions due to nonuniformity in the TiN film the MKIDs are made from.  We are currently funded by NASA to work on more uniform TiN films and are developing new materials such as PtSi that should significantly improve this issue. b: The spectral resolution is currently limited by a combination of the white noise from the HEMT amplifier, TLS noise in the resonators~\citep{2008ApPhL..92u2504G}, and geometric effects inside the MKID inductor~\citep{Mazin:2012kl}.  We are currently funded by NASA to improve the spectral resolution of our detectors, and are pursuing an all around approach of going to lower noise HEMT amplifiers and optimizing resonator geometry to improve performance.  c: The MKID QE is currently limited by the raw absorption of photons into the TiN inductor.  Using NASA funding, we are looking at designs that absorb photons in black absorbers like vertically aligned carbon nanotubes (VACNTs)~\citep{Yang:2008fl}.  This approach has the potential to give nearly unity QE across our entire wavelength band. d: The wavelength range is limited by the sky background rate in the IR and the difficulty of passing UV while simultaneously blocking thermal IR.}
\label{tab:perf}
\vspace{-0.2in}
\end{table*}

\section{Detailed Science Case}

\subsection{Compact Binaries}
\noindent\emph{By Paul Szypryt}

The exquisite time resolution, broad spectral coverage, and moderate energy resolution will make KRAKENS a valuable instrument for the study of the shortest period compact binary systems known. KRAKENS will be able to provide high signal-noise, low-resolution spectroscopy where only photometry has previously been possible. With an energy resolution of $\sim 25$, KRAKENS will be able to disentangle the physical components of an eclipsing binary (primary, secondary, disk, and bright spot) through observation during ingress and egress. With a time resolution of $\sim 1\mu s$, KRAKENS will be able to search for time-variability over a large frequency range, from lower frequency orbital periods to higher frequency quasi-periodic oscillations (QPOs) and super-orbital periods.  The time resolution will also allow KRAKENS to very finely sample the ingress and egress of an eclipsing system, providing clues for calculating important parameters such as the locations of poles in magnetic white dwarf systems. Because KRAKENS is a photon-counting instrument, it will have the additional advantage of being able to choose a pseudo-exposure time in software that maximizes the SNR, whereas in traditional CCDs the exposure time needs to be set during the observations.

\begin{figure}
\begin{center}
\includegraphics[width=\columnwidth]{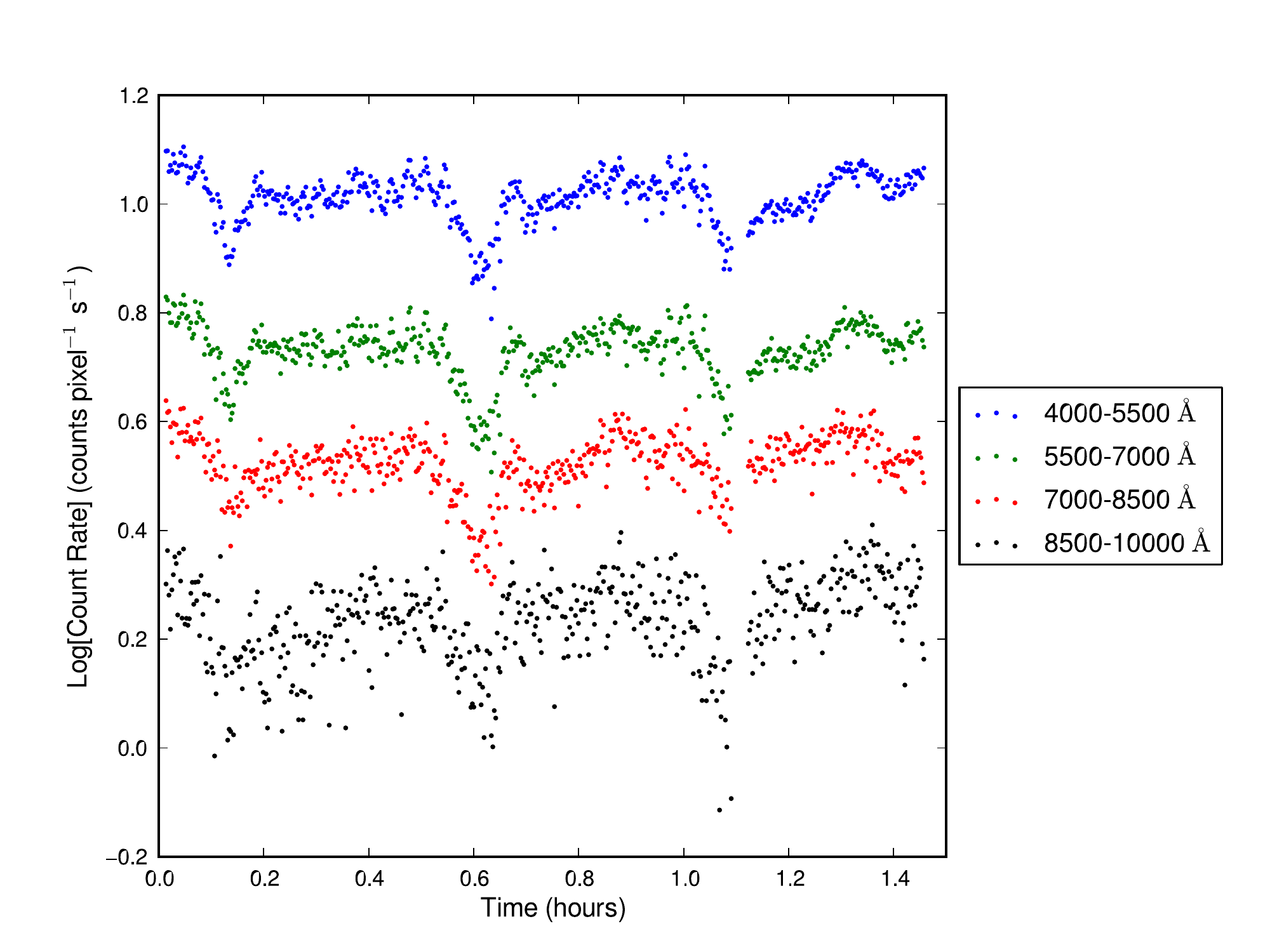}
\end{center}
\vspace{-.1in}
\caption{\footnotesize Light curves of SDSS J0926+3624 in the blue ($4000-5500$~\AA), green ($5500-7000$~\AA), red ($7000-8500$ \AA), and infrared ($8500-10000$~\AA) bands, obtained using aperture photometry. Reprinted from \citet{2014MNRAS.tmp..307S}.}  
\label{fig:J0926_Lightcurve}
\vspace{-.2in}
\end{figure}

Compact binary observations using MKIDs have been shown to be successful. In December 2012, ARCONS was used at Palomar to observe the ultracompact eclipsing AM CVn system, SDSS J0926+3624, and the light curves are shown in Figure~\ref{fig:J0926_Lightcurve}. We measured, for the first time in this object, an orbital period rate of change of $(3.07 \pm 0.56) \times 10^{-13}$~\citep{2014MNRAS.tmp..307S}. We also used the high intrinsic time resolution of MKIDs to search for QPOs to much higher frequencies than possible with other instruments, and we used the energy resolution to search for spectral variability during eclipse. KRAKENS will be able to do the same compact binary science as ARCONS, but to higher magnitudes and accuracy.  The increased throughput will allow KRAKENS to observe similar 19th magnitude objects in 1s with a SNR $\sim 100$. The increased energy resolution will allow for better separation of the various components of a binary system. In addition, the larger FOV will allow for a higher number of targets that can be studied using relative photometry.

\subsection{Exoplanet Transits}
\noindent\emph{By Gautam Vasisht, Seth Meeker, and Ben Mazin}

Due to their aligned geometries, transiting planets allow measurements of physical and atmospheric properties that are otherwise inaccessible in non-transiting systems. The transit geometry allows planetary measurements with no need to resolve the planet-star pair. The depth determines the size ratio of the planet relative to that of the star, $(R_p/R_s)$. Crucially, the atmosphere becomes accessible. It appears as opaque at certain wavelengths and transparent at others, i.e., $(R_p/R_s)(\lambda)$ carries the information on atmospheric absorbers.  

\begin{figure}%[ht]
 \begin{center}
 \includegraphics[width=1.0\columnwidth]{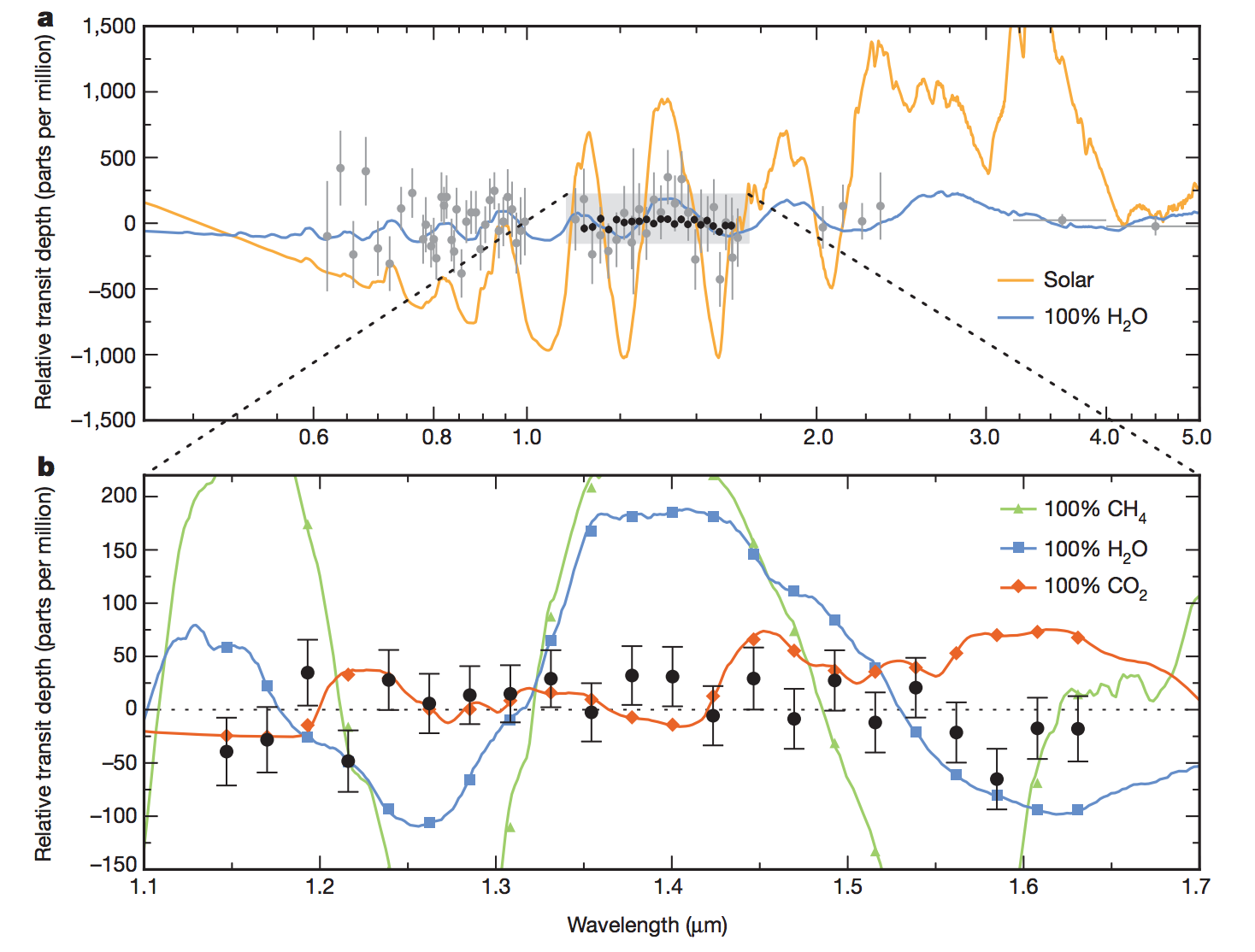}
 \end{center}
\vspace{-.1in}
 \caption{\footnotesize Transmission spectrum measurement of GJ 1214b from~\citet{2014Natur.505...69K}. They compare HST data against model spectra for atmospheres of various, notional compositions, finding the data to be consistent with a flat, cloudy spectrum. This figure exemplifies the importance of probing a broad wavelength range simultaneously as these models show strong atmospheric features across the entire KRAKENS bandwidth.}
\label{fig:transitspec}
\vspace{-.2in}
\end{figure}

Recent observations of HD 189733~\citep{2013MNRAS.432.2917P}, GJ 436b and GJ 1214b~\citep{2014Natur.505...66K,2014Natur.505...69K} in the optical and infrared all show evidence for varying amounts of clouds and/or hazes.  However, current ground and space-based observations of exoplanet transits are generally limited in the simultaneous wavelength coverage achieved and derivation of a complete spectrum often involves the combination of data taken during different orbital cycles and an assumption of non-variability which may not hold for many stars.  KRAKENS can simultaneously observe the wavelength-dependent transit depth between 350 nm and 1.35 microns.  These observations over three octaves in wavelength will be a powerful constraint of the atmospheric composition, particularly the existence of hazes and clouds as shown in Figure~\ref{fig:transitspec}. Observations of exoplanet transits are obviously highly time constrained and while these observations do not require the high time resolution of the MKIDS detector, the ability to flexibly schedule observations using K1DM3 will be highly advantageous to transit observations, and the time resolution may enable measurement of transit timing variations~\citep{2008ApJ...688..636N}.

\begin{figure}%[ht]
 \begin{center}
 \includegraphics[width=1.0\columnwidth]{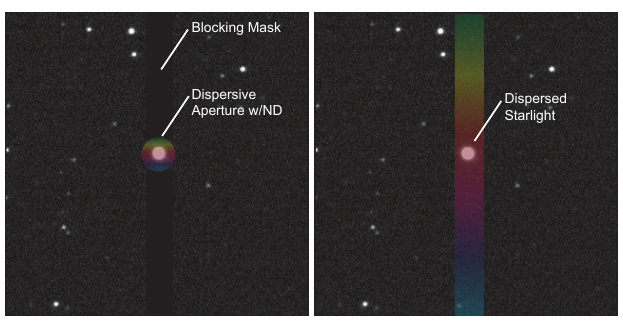}
 \end{center}
\vspace{-.1in}
 \caption{\footnotesize A new proposed transit spectroscopy technique for KRAKENS.  A wavelength dispersive zone is placed at the center of an otherwise occulting bar in the focal plane (left), allowing just the starlight to be dispersed (right), while preserving the rest of the field unchanged for atmospheric and sky monitoring and subtraction.  This will likely require slicing, collimating, dispersing, and then recombining the light in the fore module.  Further study of this concept is underway.}
\label{fig:tt}
\vspace{-.2in}
\end{figure}

Ground-based transit spectroscopy is limited by systematics due our inability to remove the effect of the Earth's atmosphere on the transmitted light and subtract the sky background.  We will consider a special exoplanet fore module for KRAKENS that will pick off and disperse the light from a central aperture (that will contain the bright star hosting the transit) and block the light in a the rest of a vertical strip (where the dispersed light will land), but pass the rest of the field unaltered.  

The effect of the this would be to allow observations of stars down to 9$^{th}$ magnitude, while optimizing the SNR of the much fainter (undispersed, unattenuated) stars in the field to serve as atmosphere and sky monitors.  This new approach, shown in Figure~\ref{fig:tt}, combined with with the MKID's intrinsic spectral resolution (for the reference stars), near-IR coverage, and high time resolution to better track fast changes in sky background and atmospheric transmission, should make it the premier ground-based instrument for transit spectroscopy.  Simulations of this mode are currently being performed at UCSB.

\subsection{Lensing Clusters}
\noindent\emph{By Brian Siana}

As mass models of galaxy clusters have become more precise, large amounts of time on space and ground-based telescopes have been allocated to study strongly lensed (and therefore highly magnified) galaxies behind lensing clusters.  This has pushed the redshift frontier to $z{\sim}$10--11 (\emph{e.g.}~\citet{Coe:2013ko}) and allowed detection of galaxies orders of magnitude fainter than previous studies at the peak epoch of star formation ($z{\sim}$2,~\citet{Alavi:2014hs}). 

Though these recent studies have been very compelling, the next generation of studies is particularly exciting. These will provide constraints on the dark energy equation of state ~\citep{Jullo:2010hv}, searches for low mass dark sub-structure in clusters, and resolved spectroscopy of lensed galaxies on physical scales smaller than 100 pc.  

Future studies, such as following up Hubble Frontier Fields, will require more precise lensing models.  This will require spectroscopic redshifts of hundreds of relatively low mass cluster members, line-of-sight galaxies and/or structures~\citep{2014MNRAS.445.3581D}, and dozens of multiply imaged background galaxies.  In every case, the spectroscopic redshift need only be precise to a few percent, and can be easily provided by identification of the Balmer or 4000 \AA~break in the KRAKENS wavelength coverage from $0 < z < 2.5$ and the Lyman break at higher redshifts.  KRAKENS will be able to obtain hundreds of redshifts simultaneously without worry of slit or fiber conflicts in these densely populated fields, and will do an especially good job on extended low surface brightness arcs from lenses. Though there are other integral field spectrographs  (VLT/MUSE), KRAKENS has at least a factor of two higher throughput, and a much larger wavelength (or redshift) range.  This means KRAKENS can obtain obtain accurate redshifts from continuum spectra for hundreds of cluster members in 1/6th the time that VLT/MUSE would require. Alternatively, in the same amount of time, KRAKENS can measure continuum photometric redshifts for galaxies one magnitude fainter than VLT/MUSE. 

While the accurate photometric redshifts are useful for characterizing the lenses, the detailed spectral energy distributions over the broad KRAKENS wavelength range will allow more detailed investigations.  For example, studies of the cluster galaxies' stellar population ages as a function of cluster-centric radius to test models of quenching, and rest-frame ultraviolet spectral slopes of lensed galaxies at $z > 1.5$ to measure dust extinction levels and ultimately determine the total star formation rates. 

%Jones et al. 2013 ApJ 765, 48

\subsection{Gravitational Wave Counterparts}
\noindent\emph{By Jason X. Prochaska}

The upgrade of the Laser Interferometer Gravitational Wave
Observatory (LIGO) to Advanced LIGO has entered its commissioning
phase. With full sensitivity expected by 2018, 
the gravitational wave (GW) community is
cautiously optimistic for the first detections of GW events from
compact mergers at cosmological distances. 
Alerts for putative events will be broadcast widely and teams of
astronomers are poised to search the large patches on the sky 
corresponding to the GW `localization'
(hundreds to thousands of square degrees) 
for electromagnetic (EM) counterparts~\citep{2014ApJ...789L...5K}.
The discovery of an EM counterpart would offer the opportunity to
resolve the origin of the GW event~\citep{2011ApJ...739...99N},
and a source redshift would yield
the energetics and enable unique tests of cosmology~\citep{1986Natur.323..310S,2006PhRvD..74f3006D}.

Given the relatively modest field-of-view of KRAKENS, its
participation in this arena would be to vette candidate EM
counterparts drawn from the LIGO error box by other ground-based
imagers.  When coupled to the K1DM3 tertiary system (expected
commissioning by early 2017), the KRAKENS camera could rapidly
generate an SED for $\approx 10$ faint sources ($R \gtrsim 25$)
per hour. On its own, this dataset may confirm the counterpart.  At the
least, KRAKENS would further refine the candidate list and identify
the most promising targets for deeper spectroscopy.

Regarding the requirements for KRAKENS, the vetting of candidate
counterparts would benefit from higher spectral resolution and
(probably) higher sensitivity at redder wavelengths.  The former is
critical to rule-out sources such as variable stars, flaring AGN, solar
system phenomena, etc.  The latter follows current expectation that
the counterparts will emit preferentially at longer wavelengths~\citep{2013ApJ...775...18B}. 

\begin{figure}[t]
   \begin{center}
   \includegraphics[width=\columnwidth]{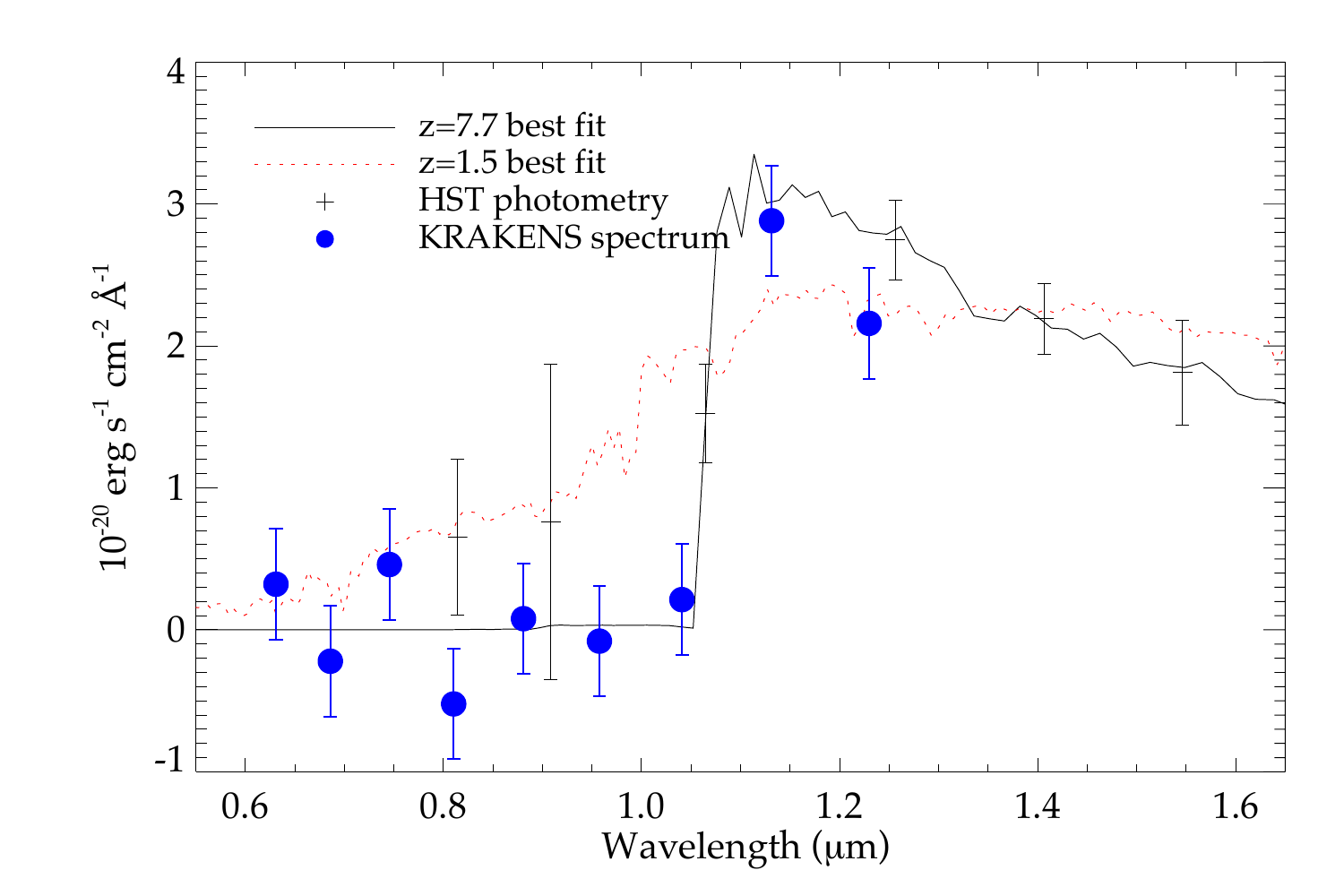}
   \caption{\footnotesize HST photometry of the candidate $z{\sim}8$ galaxy ``a611-0193" selected by~\cite{Bradley:2014gk}. The source shows a strong spectral break at ${\sim}$1 micron. Best-fit photometric templates are shown for a Lyman break galaxy at $z{\sim}8$, and a $z{\sim}1.5$ galaxy with a strong Balmer break. Deep HST data are unable to distinguish these two cases. Blue points show a simulated 5-hour KRAKENS observation for the z${\sim}$8 case. The KRAKENS spectrum is able to securely identify the redshift as z${\sim}$8, ruling out the low-$z$ solution at 6-$\sigma$ significance. This is enabled by KRAKENS's good sensitivity and improved spectral sampling compared to HST.  The simulated spectrum assumes 0.4 arcsecond resolution (i.e. with tip/tilt correction), 5 hours of on-source exposure time, a spectral resolution R=10 for the wavelength range shown, and the ``low throughput" case.}
   \label{fig:highz}
   \end{center}
\vspace{-.2in}
\end{figure}

\FloatBarrier

\subsection{Galaxies in the Epoch of Reionization}
\noindent\emph{By Tucker Jones}

The wide spectral coverage and excellent sensitivity of KRAKENS will be a powerful tool for measuring redshifts of faint galaxies. Redshifts for faint sources are typically determined from either strong emission lines, or continuum breaks seen in broad-band photometry. The latter, however, are subject to potentially catastrophic uncertainties. This is especially problematic for candidate galaxies at very high redshifts ($z > 7$) selected on the basis of a potential Lyman break. Typically these sources show no strong spectral features and are difficult to separate from $z{\sim}$1--2 galaxies and low-mass stars. Figure~\ref{fig:highz} illustrates this problem, showing that deep HST photometry is unable to distinguish between $z{\sim}1.5$ and $z{\sim}8$ as the redshift of a faint source. KRAKENS will be able to identify the correct redshift solution in a modest integration time. It will therefore enable higher fidelity surveys of the galaxy population at redshifts up to $z > 8$. Simultaneously, deep surveys with KRAKENS will provide an excellent ``training set" for photometric redshifts of faint objects which are difficult to identify with current facilities.

\subsection{Kuiper Belt Objects}
\noindent\emph{By Wesley Fraser}

\begin{figure}[t]
   \begin{center}
   \includegraphics[width=\columnwidth]{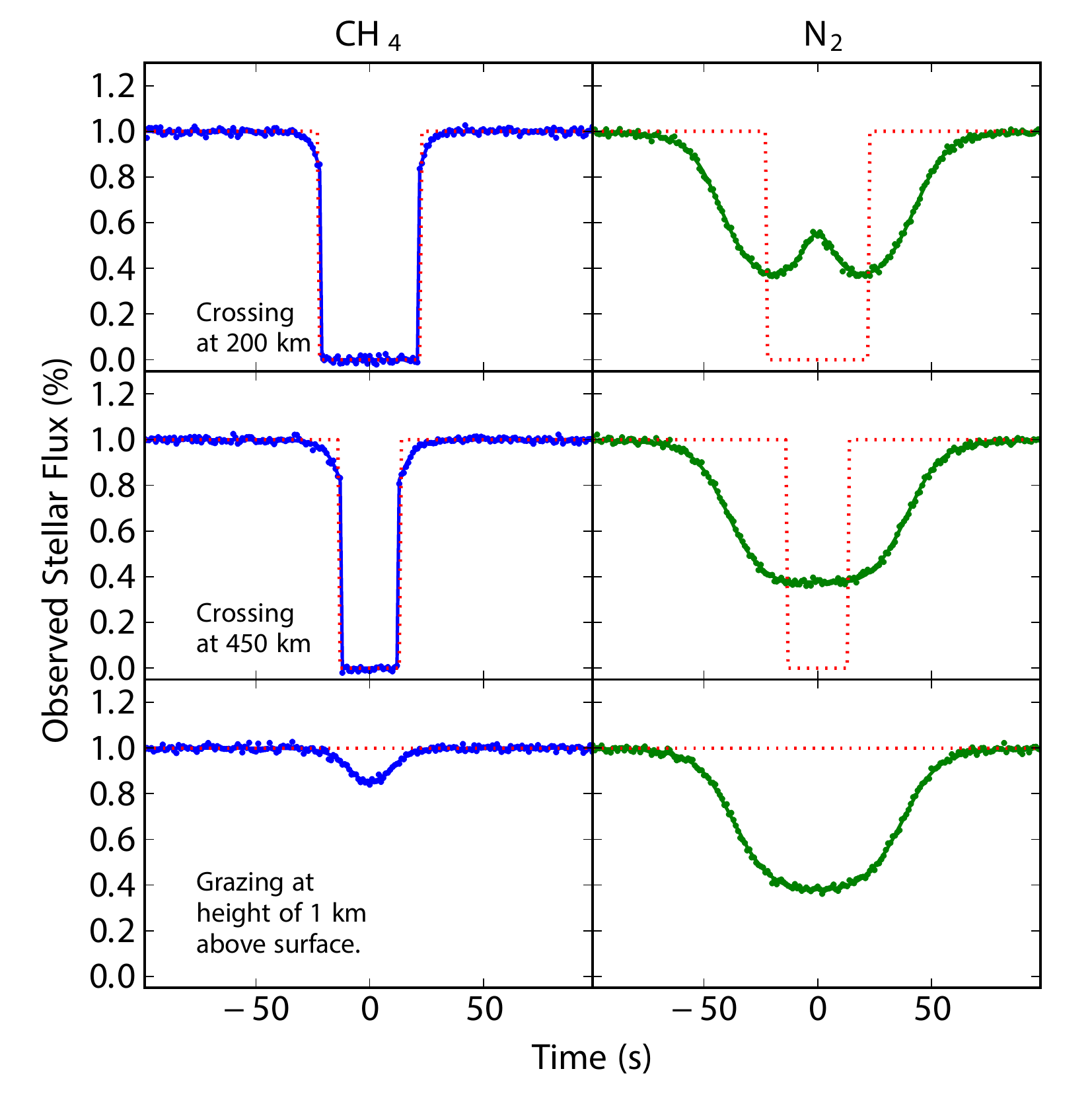}
   \caption{\footnotesize Simulated occultations by a hypothetical 1100 km diameter body with a 26 km s$^{-1}$ shadow velocity. Three occultation geometries and two atmospheric compositions are shown. The atmosphere is in sublimation equilibrium at surface temperature, 41 K, which results in surface pressures of 16 nbar and 100 microbar for the methane and nitrogen atmospheres respectively. The solid lines show the model light curves, and the red dotted curve shows that if no atmosphere were present, respectively. The points are simulated 1 second observations with a SNR=100, which is expected for an R=19 star as shown in Figure~\ref{fig:perf}. The atmospheric signatures which may include refraction central peaks (see top right) are easily detected. }
   \label{fig:occult}
   \end{center}
\vspace{-.2in}
\end{figure}

Observing stellar occultations is a very powerful technique to learn about the small-body populations of the Solar system. The observation of a stellar occultation by a planetesimal can allow the accurate determination of the occulting body's shape and size (eg., Ortiz et al. 2012). As well, barring exploratory space missions, stellar occultations present the only method by which the existence of a tenuous atmosphere about a planetesimal can be revealed (see Figure~\ref{fig:occult} and~\cite{Elliot1989}). This fact is reflected by Kuiper belt objects (KBOs); while it is expected that many KBOs will have atmospheres~\citep{2008ssbn.book..365S}, Pluto is the only KBO whose atmosphere has been detected --- the identification was made by stellar occultation.

Observations of stellar occultations have been hampered primarily by the lack of detectors capable of the $>$1 Hz photometry necessary to accurately characterize the occultation light curves, which typically last no more than 60 s (see Figure~\ref{fig:occult}). The natural time resolution of the MKIDS clearly make them an ideal detector for occultation observations. Further, compared to EMCCDs, MKIDS detectors have greater bandwidth, allowing telescopes to observe occultations of fainter, and therefore more numerous occultations than otherwise possible.

\begin{figure}[t]
   \begin{center}
   \includegraphics[width=\columnwidth]{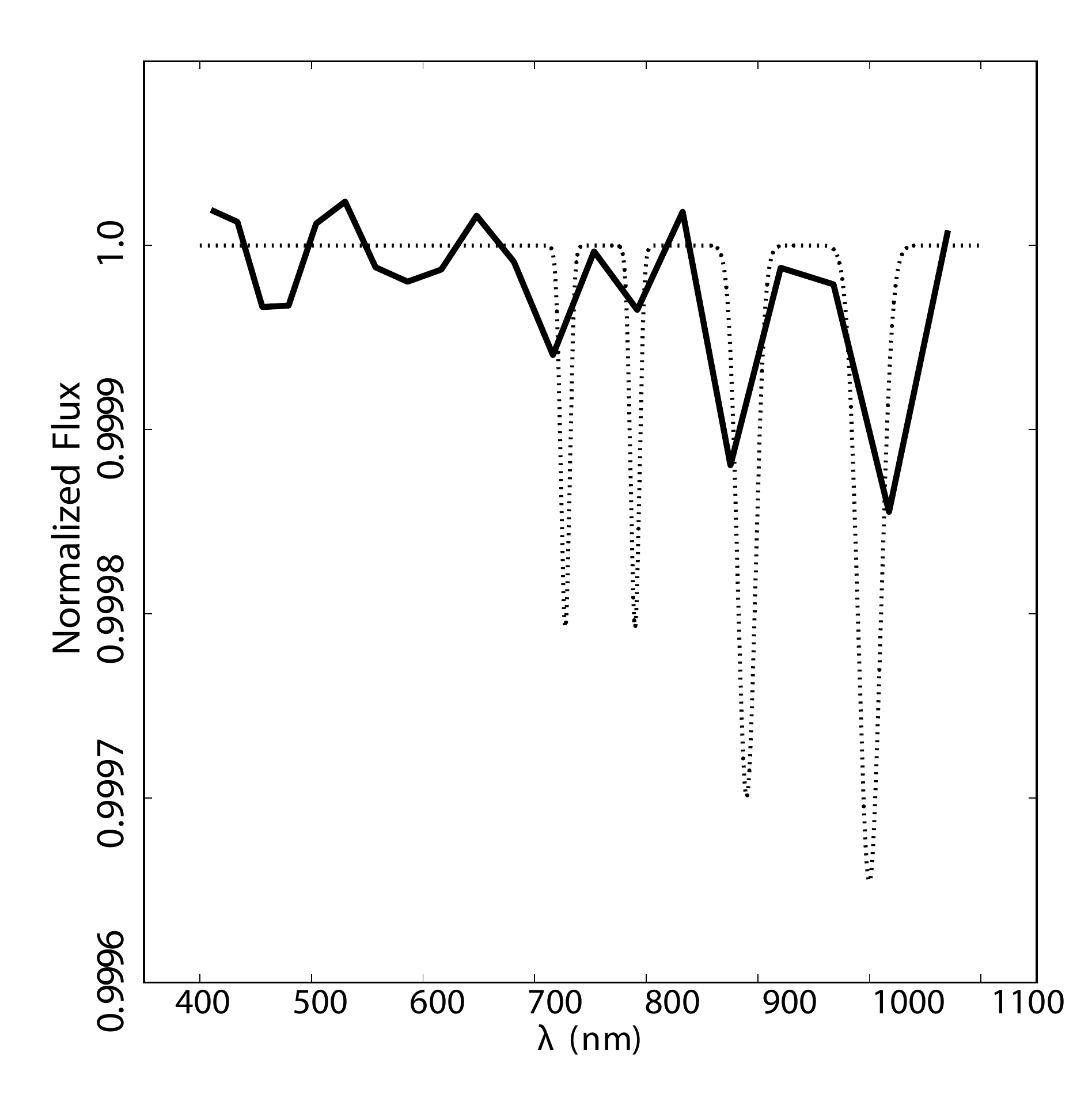}
   \caption{\footnotesize  Dotted line: Model atmospheric transmission spectrum of a pure CH$_4$ atmosphere around a Kuiper Belt Object, the same atmosphere as used in the models presented in Figure~\ref{fig:occult}. Solid Line: Observed spectrum from KRAKENS with R=20, occulting a R=16 source. The methane absorption bands present only a 0.1\% signal, but are still detectable by the MKIDS detector. }
   \label{fig:ch4}
   \end{center}
\vspace{-.2in}
\end{figure}

With a single band, KBO atmospheres can be detected, and their pressure-temperature profiles can be characterized. Little however, can be said about compositions. This is not the case with an MKIDS, which can provide course spectra of the atmosphere transmission spectrum (see Figure~\ref{fig:ch4}). With a sufficiently bright source and a defocused pupil plane image or using the grating fore module to allow higher count rates, detections of methane and nitrogen absorption bands at ${\sim}$1 micron may be possible. Such a detection is impossible with any other detector.

\begin{figure}[t]
   \begin{center}
   \includegraphics[width=\columnwidth]{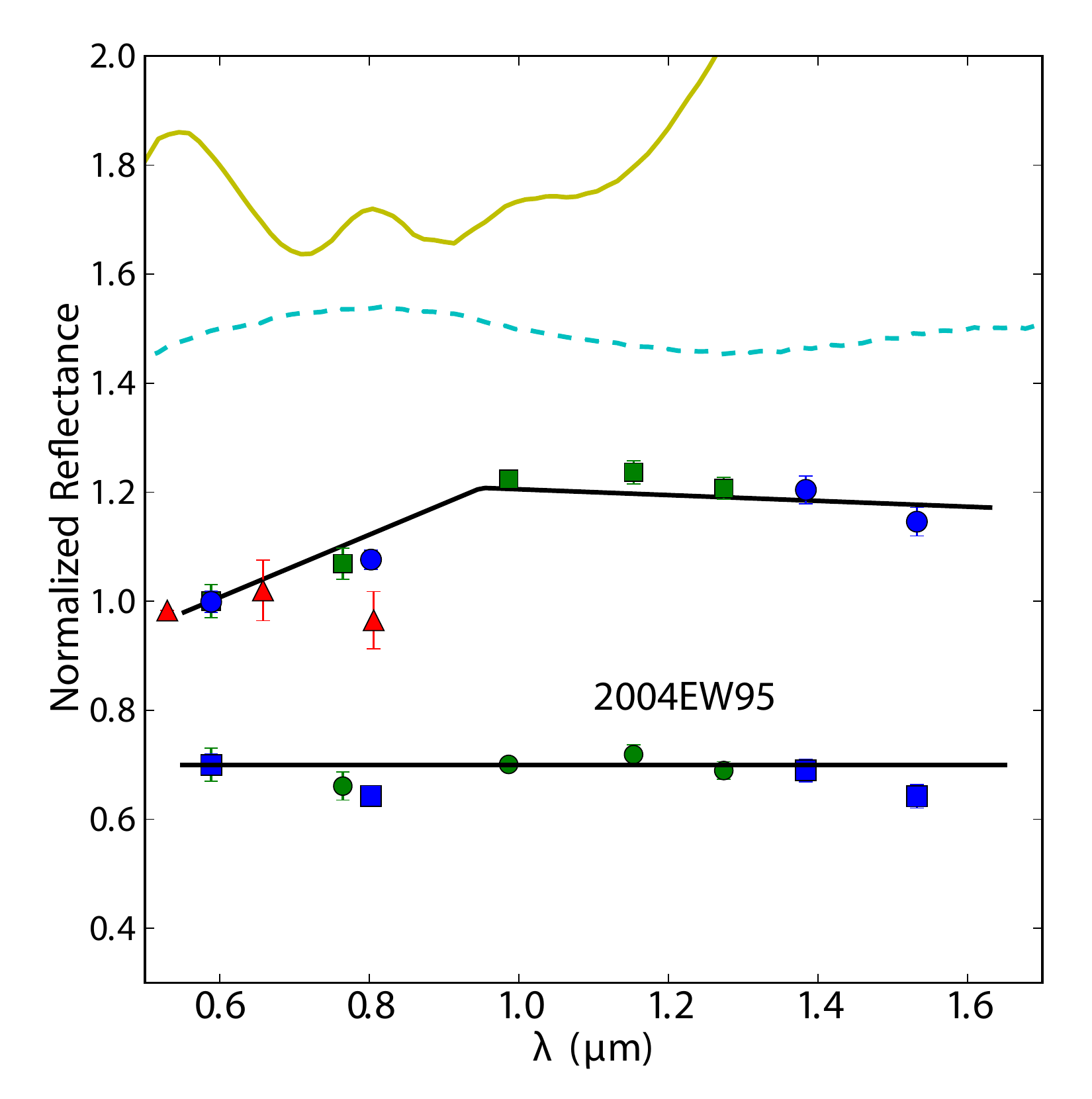}
   \caption{\footnotesize  Blue and green points: HST broad band photometry of Kuiper Belt Object 2004 EW95. Black line: linear spectral model, fit to the data. Residuals of the photometry and the model are shown, offset below the observations, and reveal the existence of a features at ${\sim}$0.7 and ${\sim}$1.1 microns. These features are consistent with silicate absorption features seen on Earth and in the asteroid belt. Example of some silicates that are consistent with the features inferred in EW95's spectrum are shown offset above for comparison (anorthite - blue, and chlorite - yellow). }
   \label{fig:kbo}
   \end{center}
\vspace{-.2in}
\end{figure}

To date, spectroscopic studies of Kuiper Belt Objects have revealed purely icy surfaces with no signatures of non-icy or silicate materials present. This is presumably a result of differentiation processes which, over the age of the Solar System, have allowed differentiation to occur, hiding dense non-icy materials below low density, purely icy mantles~\citep{McKinnon2008}. Observations by~\cite{Fraser2012} have suggested that small KBOs, with diameters D$<$200 km are no longer differentiated, and exhibit hints of silicate features in their broad-band photometry (see Figure~\ref{fig:kbo}). Currently, no detector exists with the sensitivity to get high quality spectra of such faint objects; available spectrographs observe at too high a resolution, and hence have too low effective efficiencies. Furthermore, few detectors simultaneously observe at the critical wavelength of $\sim$1 micron where many silicate features exist. The high throughput and low spectral resolution of the MKIDS detector presents the first opportunity to acquire high quality spectra on a number of small KBOs, and search for the features indicative of silicates in the Kuiper Belt. With the detection and identification of what silicates make up the bulk of KBOs, we can compare those to the silicates we see in other regions of the Solar system, and answer questions like: At what location in the primordial belt did KBOs form? How hot must the formation environment have been to result in the ice-to-rock fraction observed in KBOs? Detection of silicates has the potential to revolutionize our understanding of the early Solar System's formation, and is only currently possible with the MKIDS detectors.

\subsection{Strong Lensing}
\noindent\emph{By Tommaso Treu}

In the standard cosmological model, the matter density of the Universe is dominated by a mysterious non-relativistic particle (or particles) known as dark matter. Dark matter does not interact with light and therefore has been detected only through its gravitational effects on astronomical scale. Cosmological simulations predict that dark matter halos of galaxies should be surrounded by thousands of satellites.  Yet, the number of observed satellites of the Milky Way is much smaller, even taking into account recent discoveries.  This discrepancy implies either that satellites do not form stars and are therefore dark and invisible, or that the cold dark matter scenario is fundamentally wrong. For example, if the dark matter subhalo mass function cuts off below a certain mass, this could be an indication that dark matter is warm, instead of cold (\emph{e.g.} a keV scale sterile neutrino).

Dark satellites, if present, can be detected via the perturbation of the positions and fluxes of multiple images of background lensed galaxies. The current detection limit at HST resolution is a few $10^8$ solar masses~\citep{2010MNRAS.407..225V,2013MNRAS.436.2120N}, i.e. at the high mass end of the MW satellite mass function where the agreement between theory and observations is best. KRAKENS behind adaptive optics can make progress by exploiting the fact that lensing is achromatic. Traditionally, gravitational lenses are modeled by exploiting the conservation of surface brightness. However the energy of each photon is also conserved and therefore KRAKENS would multiply the number of constraints available for any given model with respect to traditional imaging. The idea has been proposed for integral field spectroscopy (Melling, Treu \& Marshall 2006, unpublished) and for ALMA interferometry~\citep{2013ApJ...767....9H} and shown to be very powerful. KRAKENS would be a cost effective implementation of the idea by providing spectroscopic information at the cost of imaging.

\subsection{X-ray Binaries}
\noindent\emph{By Kieran O'Brien}

LMXBs are interacting binaries containing a low mass donor star transferring matter onto a neutron star (NS) or a black hole (BH). They offer us the opportunity to study both the accretion physics and the properties of the compact stellar remnants. LMXBs are multiwavelength variable sources emitting from hard X-rays to radio, and displaying variability in time-scales from milliseconds to years. During the last decades, X-ray observations of LMXBs at high-time resolution have revealed a rich phenomenology, allowing us to probe the inner regions around neutron stars and black holes (see \emph{e.g.} \citet{2006csxs.book...39V}). KRAKENS will enable us to gain insights into a new region of parameter space through correlated X-ray and OIR observations, which holds exciting potential for the study of the physics of accretion and evolution of compact objects in X-ray binaries.

\noindent {\bf 1000 to 0.1 seconds: Reprocessing and Echo-Mapping}

\begin{figure}
\begin{center}
\includegraphics[width=0.8\columnwidth]{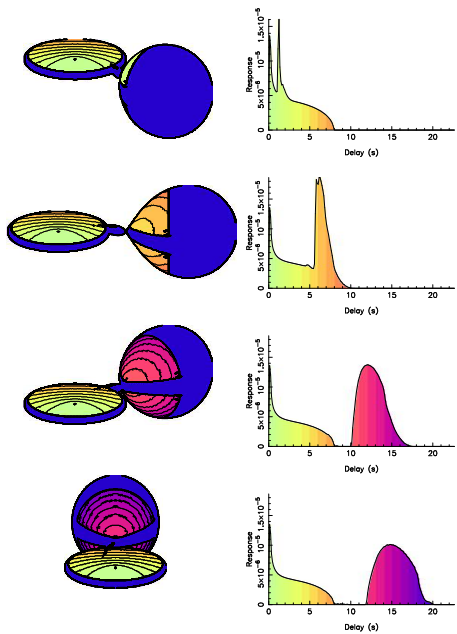}
\end{center}
\caption{\footnotesize Left column: schematic view showing the irradiated regions of a typical LMXB (based on the binary parameters of Scorpius~X-1) as a function of binary phase (top to bottom). Right column, the associated time delay transfer function which describes the observed time delay between the directly observed X-ray flux variability and the resulting lower energy (opt/IR) flux changes (from \citet{2002MNRAS.334..426O}).} 
\label{fig:transfer}
\end{figure}

When X-ray reprocessing takes place, the OIR emission seen by a distant observer should be delayed in time relative to the X-rays due to light travel times ($\sim$few seconds) between the X-ray source and the reprocessing sites within the binary system. The time-delayed and distorted optical echoes directly measure the locations and sizes of reprocessing sites (see Figure~\ref{fig:transfer} from~\citet{2002MNRAS.334..426O}). Time-delay transfer functions can be computed by studying the correlated variability in X-ray and OIR. This has been carried out for a number of different objects. For instance, several simultaneous X-ray/optical bursts have been observed to originate from reprocessing (\emph{e.g.}, \citet{pedersen82}). These observations used broadband optical photometry, which is sensitive to reprocessing regions within the accretion disc. More recently, \citet{2007MNRAS.379.1637M} have used RXTE data and narrow band optical observations of a blend of emission lines arising from reprocessing on the companion star (see \emph{e.g.} \citet{2002ApJ...568..273S}). Observations have also been made on neutron star systems (Sco X-1 and 4U 1636-536) at high time resolution ($\sim$0.1-2 s) during flaring and thermonuclear burst episodes, showing that the delays associated with emission lines and continuum are different. The former is consistent with the companion star and allow us to map the binary orbit and constrain dynamical parameters like the orbital inclination (Figure~\ref{fig:transfer}), whereas the latter seems related with reprocessing within an accretion disc with a radial temperature profile and allow us to study the accretion geometry.  KRAKENS will extend these kind of observations to fainter targets and faster time domain signals.

\noindent {\bf 1 to 0.001 seconds: Variability from Jets}

\begin{figure}
\begin{center}
\includegraphics[width=1.0\columnwidth]{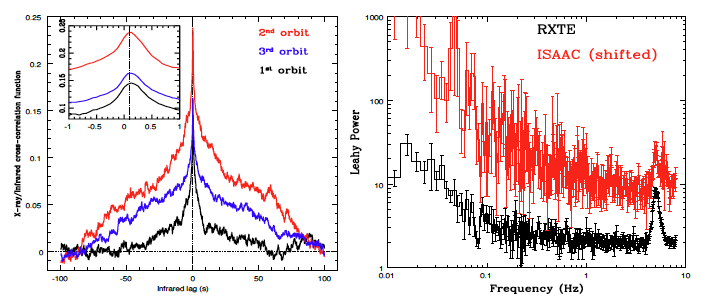}
\end{center}
\caption{\footnotesize Left panel: Cross-correlation function between the RXTE and ISAAC K-band light curves for GX~339-4~\citep{2005ApJ...629..403C}. The symmetric pattern and the small delay (0.1 s) indicate a jet origin for the IR variability and allowed us to estimate a relativistic jet speed. This result represents the first detection of sub-second jet variability. Right panel: Power density spectra for RXTE and ISAAC (shifted vertically for clarity) light-curves for GX 339-4 (Casella et al., in prep). A 5 Hz, type-B quasi-periodic oscillation characteristic of accretion states where relativistic jet ejections are observed is present in both data sets~\citep{2005ApJ...629..403C,2008MNRAS.383.1089S}. This result represents the first detection of this timing feature in the IR, suggesting a jet origin for it.} 
\label{fig:gx339}
\end{figure}

Spectral studies have shown in the recent years that the OIR emission can include a substantial contribution from the relativistic jet, in both BH and NS X-ray binaries~\citep{2004MNRAS.351..253M}. Surprisingly complex cross-correlation functions (CCFs) were also discovered in BHs, with X-rays partially lagging optical emission (\emph{e.g.}~\cite{2002A&A...391..225S,2006ApJ...648.1156H,2008ApJ...682L..45D,2008MNRAS.390L..29G}). The most probable physical explanation for this unusual behavior involves a key role played by a powerful relativistic jet~\citep{2004MNRAS.351..253M}. Using ISAAC data, a clear correlation was found between IR and X-ray variability in the BH system GX339-4 (\citet{2005ApJ...629..403C}; see right panel in Figure~\ref{fig:gx339}). Simple considerations and calculations permitted the identification of a jet origin for the observed IR variability, resulting in the first unambiguous evidence for sub-second variability in an X-ray binary jet. This technique provides access to phenomena that take place at the base of the jet, very close to the compact object, where fundamental physics like particle acceleration, shock formation, and magnetic reconnection can be studied in detail. 

The above methods have great potential, but only now are starting to give us new insights. KRAKENS offers a truly unique capability for such observations, allowing spectroscopic studies on the fastest timescales. The read-noise free detector with near-IR coverage will allow us to study the correlated variability at millisecond timescales with a signal-to-noise similar to that achieved on an 8-m telescope with standard instrumentation (c.f. 50~e-/pix/DIT for ISAAC/VLT). These observations of X-ray binaries will enable us to understand the origin of the OIR emission in these systems and constrain the accretion geometry and jet physics.

\subsection{Magnetars}
\noindent\emph{By Shriharsh Tendulkar}

Magnetars\footnote{For recent reviews, see \citet{mereghetti2013, olausen2014}.} are very young neutron stars (age $<10$\,kyr) with long rotation periods. Unlike radio pulsars, the intense magnetic field of magnetars ($B\sim10^{15}\,$G) serves as the  primary energy reservoir of magnetars whose giant flares are the most luminous ($L\sim10^{47}\,\mathrm{ergs\,s^{-1}}$) events in the Universe and could be seen as short GRBs from other galaxies~\citep{hurley2011}. While magnetars tend to be well-studied in X-rays and radio, our understanding of these enigmatic objects is limited by (a) the few OIR counterparts known (6 out of 25 known magnetars have confirmed OIR counterparts, and (b) the scarcity large-telescope instruments capable of time-resolved observations. There is little consensus on the source of OIR emission from magnetars although the OIR luminosity is many times higher than expected from the thermal X-ray spectra~\citep{israel2003,kaplan2011}. With its timing and spectral resolution, KRAKENS will be able to (a) indisputably identify OIR counterparts based on pulsations (instead of colors or long-term flux variations), (b) measure spectral variation of pulse fraction and (c) measure polarizations (with the polarization fore module) to understand the source of the OIR emission from magnetars.

There are two central ideas for OIR emission: (a) the magnetospheric models and (b) reprocessed emission from a fall-back disk. Each of the models in Table~\ref{tab:polarization_signal} predict different spectral slopes ($\alpha$, where the spectral flux density $F_\nu \propto \nu^{\alpha}$) for the pulsed emission, different pulse fractions and polarization signals in the OIR bands. This implies the measurement of pulsations in multiple OIR bands is an effective method for distinguishing between these models. 

The evidence for the presence of a fall-back disk is conflicting. The optical to mid-IR spectrum of 4U\,0142+61 and 1E\,2259+586 shows a thermal bump from a heated disk~\citep{wang2006,kaplan2009}. However, the fall-back disk model is contradicted by OIR variability that had no corresponding X-ray variability~\citep{durant2006} and the 27\% pulse fraction in optical emission being higher than the pulse fraction in its X-ray emission~\citep{kern2002}.  The caveat in comparison between the IR pulse fraction and the X-ray pulse fraction is that they were not measured simultaneously, and hence may represent a change in emission state. This study can be systematically undertaken with coordinated KRAKENS and Swift-XRT/NuSTAR observations. No other magnetars have thermal bumps in the IR bands suggestive of a heated disk. 

\begin{table*}
\small
\begin{center}
\begin{tabular}{lcccl}
\hline
\hline	
Mechanism & Lin. Pol. & Cir. Pol. & $\alpha$ & Reference \\
\hline
Disk Blackbody (Rayleigh Jeans Tail)             & 0\% & 0\% & +2.0 &  \\ 
Coherent Emission from Plasma Instabilities      & few--10 \% & few--10\% & +0.33 & Eichler et al., 2002 \\
Synchrotron Emission from Relativistic Electrons & Strong & Strong & +2.5 & Beloborodov \& Thompson, 2007\\
Curvature Radiation                              & Strong & Weak & +0.33 & Beloborodov \& Thompson, 2007\\
Disk Reprocessing                                & $\sim0$\% & $\sim0$\% & $-$0.5 & Ertan \& Cheng 2004\\
QED emission                                     & Strong & -- & 0.0 & Heyl \& Hernquist, 2004\\
\hline
\hline	
\end{tabular}
\end{center}
\vspace{-.1in}
\caption{\footnotesize Expected signal for various Magnetar emission mechanisms. $\alpha$ is the spectral slope, where the spectral flux density is $F_\nu \propto \nu^{\alpha}$ for the pulsed emission.}
\label{tab:polarization_signal}
\vspace{-.2in}
\end{table*}

\subsection{Optical Pulsars}
\noindent\emph{By Matthew Strader}

\begin{figure*}
\begin{center}
\includegraphics[width=2.0\columnwidth]{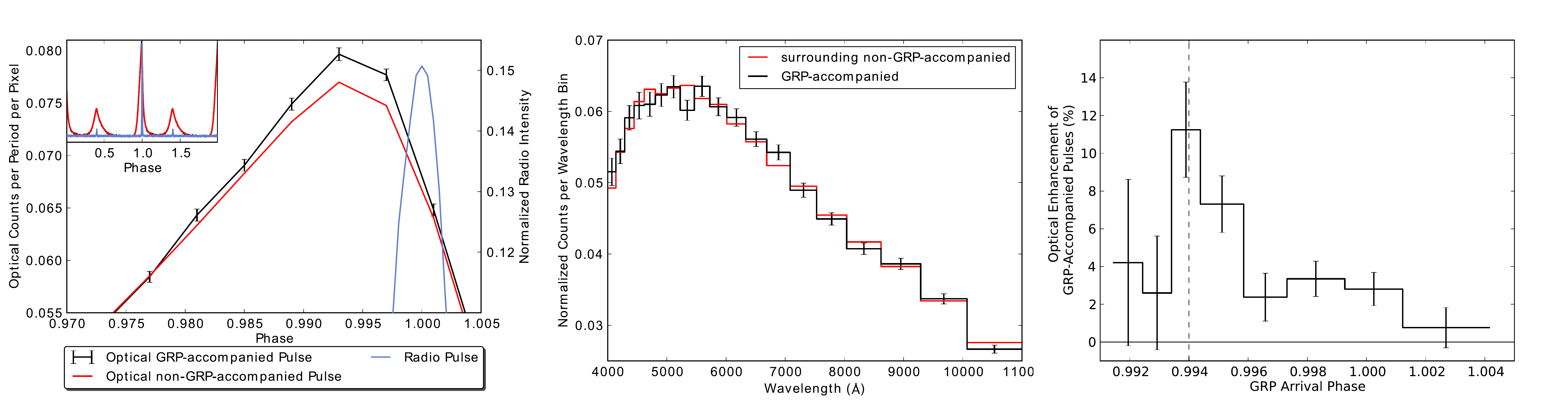}
\end{center}
\vspace{-.1in}
\caption{Left: The average optical pulse profile of the Crab Pulsar for pulses accompanied by giant radio pulses (GRPs) (black) and for pulses surrounding (within 40 periods of) GRP-accompanied pulses (red).  The profile for GRP-accompanied pulses shows about 3\% more optical counts at the peak of the main pulse (inset) than surrounding pulses.  Center: The spectrum of photons arriving during the main peak (3 phase bins with highest counts) for GRP-accompanied pulses (black) and surrounding pulses (red) normalized to have the same integrated flux.  There does not appear to be significant spectral differences between enhanced GRP-accompanied pulses and non-GRP-accompanied pulses.  Right: The optical enhancement as a function of GRP arrival phase.  This data is not consistent with a constant value at the 3.3 $\sigma$ level.  The enhancement seems to be higher for GRPs that arrive coincident with the optical pulse, shown as a dashed vertical line.}
\label{fig:crab}
\vspace{-.15in}
\end{figure*}

Despite nearly half a century of intense study, many aspects of pulsars remain enigmatic.  For instance, while reasonable models exist that can explain some observational aspects~\citep{Cheng:2000in,kern03,dyks04,Harding:2008bn}, such as the source of the $\gamma$-ray emission, there is no coherent picture that ties together a pulsar's multi-wavelength emission with other key parameters like age, spin down luminosity, radius, and surface temperature~\citep{zsk+02,szk+06,km06,2011ApJ...743...38D}.  The phase-resolved spectral information KRAKENS would provide could contribute vital clues to the determination of the optical emission mechanism.  Most theorists believe the optical emission comes from synchrotron radiation from a power law distribution of particles in the pulsar magnetosphere.  Spectral signatures, like spectral breaks from synchrotron self-absorption and cyclotron lines could, if detected, be direct probes of magnetic field strength and particle densities in the optical emitting region~\citep{1998A&A...332L..37M,Gil:2001dq}.  Combining data from KRAKENS with observations in the radio, X-ray, and $\gamma$-ray could further assist theorists in their attempts to synthesize a complete model of pulsar emission, for example~\citet{Harding:2008bn} efforts on the Crab Pulsar.  A polarimeter module would make KRAKENS pulsar data even more complete.  Data obtained with MKIDs have already contributed to the study of pulsar emission. Observations by ARCONS on the Crab pulsar, along with simultaneous radio observations by the GBT, led us to find a new temporal relationship between the arrival time of giant radio pulses and the brightness of optical pulses (see Figure~\ref{fig:crab}) \citep{2013ApJ...779L..12S}.  

KRAKENS can also be a powerful tool for discovering optical pulsars.  Optical pulses from only five rotation powered pulsars (RPPs) have been detected, and the number of definitive optical detections of any type of isolated neutron star (pulsar or otherwise) is below
20~\citep{Mignani:2011gl}. Few optical counterparts have been identified due to multiple reasons.  One reason is that they are generally faint.  Another is that the dominant astronomical detector technology, CCDs, are poor sensors for the fast observations needed to reconstruct the optical pulse profile of a pulsar.  A third is that many pulsars lie in the galactic plane, where extinction is a serious issue, so the lack of near-IR photon counters has limited progress.  The capabilities of KRAKENS addresses all of these issues with its high sensitivity (See Figure~\ref{fig:perf}), high time resolution, and wide spectral bandwidth extending into the near-IR.  In addition, many pulsars are located in crowded fields, so a number of possible faint counterparts can be identified.  KRAKENS can check all the sources in its field of view for pulsations with a known period to determine which is the pulsar.  Searches for optical counterparts for isolated millisecond pulsars (MSPs) have been unsuccessful~\citep{Sutaria:2003kl,Koptsevich:2003fx,Mignani:2004kj}.  There have been detections of optical emission from MSPs like IGR J00291+5934~\citep{DAvanzo:2007fe}, but this light has usually been attributed to a binary companion.  If pulsations are detected it could begin a new era of MSP timing for gravitational wave detection. The optical/near-IR does not suffer from interstellar dispersion or scintillation~\citep{You:2007kt}, which are likely the dominant source of timing noise in the radio. The different systematics could yield superior pulsar timing, enhancing or possibly enabling decection of the stochastic gravity wave background.

\subsection{Protostars}
\noindent\emph{By Kevin France}

\begin{figure}%[t]
   \begin{center}
   \includegraphics[width=\columnwidth]{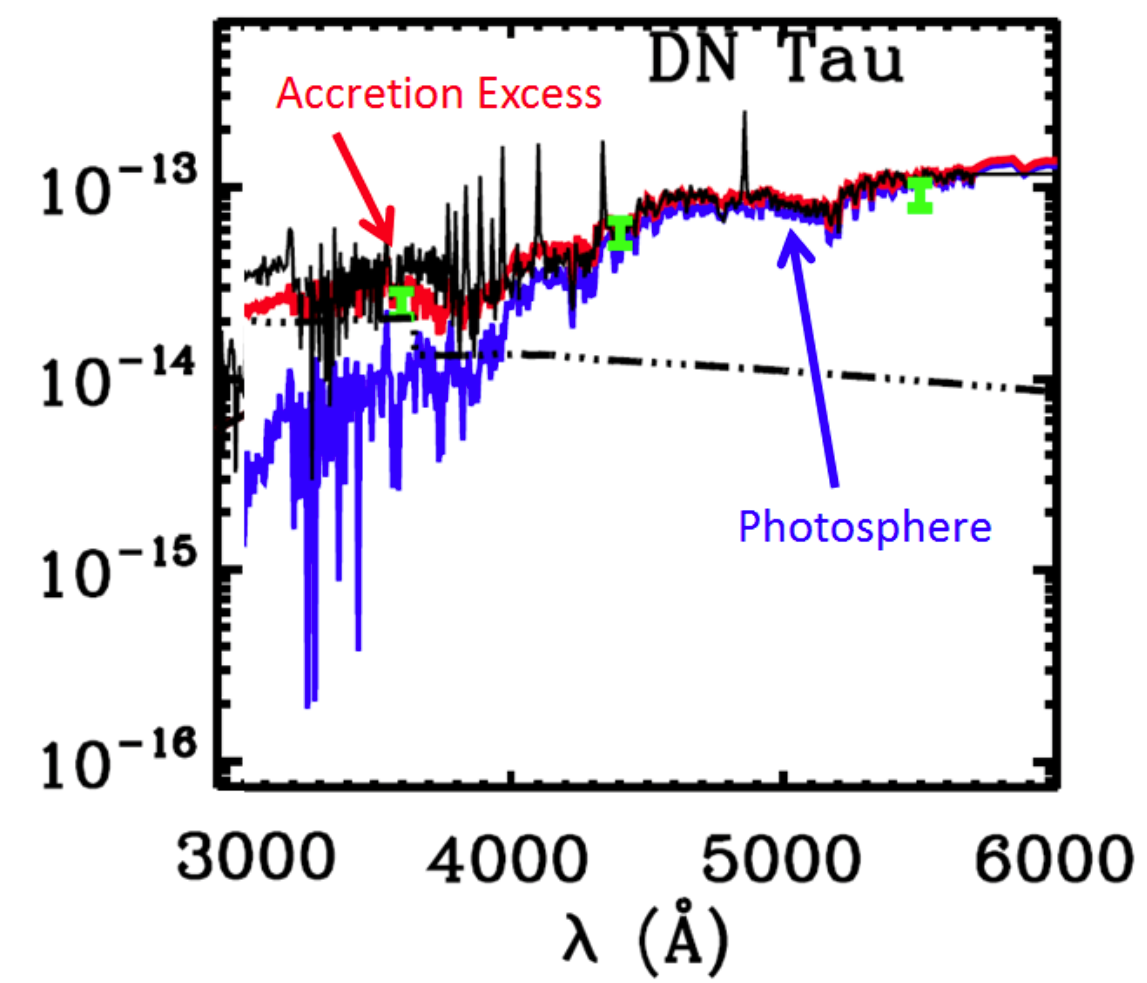}
   \caption{\footnotesize Luminosity excess in the U band is indicative of mass accretion of the circumstellar gas envelope around ${\sim}$solar mass protostars (figure adapted from \cite{Ingleby:2013hf}).  Low-resolution integral field spectroscopy is ideal for quantifying mass accretion rates of statistical samples of young stars across a range of star forming environments.}
   \label{fig:accr}
   \end{center}
\vspace{-.3in}
\end{figure}

Accretion of a circumstellar envelope onto the central object in protostellar systems is a fundamental process to the current star and planet formation paradigm~\citep{Calvet:1998ug}.  In this picture, the circumstellar material is coupled to the stellar magnetic field.  This disk gas is funneled onto the protostar along field lines, leading to strong emission from so-called accretion shocks when this infalling material piles up at the poles of the protostellar photosphere~\citep{KONIGL:1991vc}.  This process both regulates the stellar angular momentum dissipation during the formation of low-to-intermediate mass stars ($\textrm{M}_* < {\sim} 2 \textrm{M}_{\odot}$ ) and produces the high-energy radiation that drives the chemistry of the protoplanetary medium and the dispersal of the natal gas and disk, thereby setting the ultimate timescale for the formation of gas giant planets and rocky cores.   The two most widely used methods for quantifying the mass accretion rate, M$_{acc}$ [typically in units of $\textrm{M}_{\odot}$ yr$^{-1}$],  are high-resolution H spectroscopy and low-resolution U band spectroscopy or narrow-band imaging.  Accretion is characterized by U band emission in excess of more evolved, non-accreting sources, as shown in Figure~\ref{fig:accr}, owing to Balmer continuum emission originating in the shock region.  

KRAKENS is the ideal instrument to expand the study of protostellar mass accretion to diverse star-forming environments both in (d $\le$ 500 pc) and beyond the nearby (d $>$ 1 kpc) Galactic neighborhood.  Low-resolution intergral field spectroscopy at Keck provides both the 0.5 -- 0.85 $\mu$m spectrum with which to estimate the spectral type and luminosity of the protostar and the U band flux necessary to measure the excess accretion luminosity, and thus M$_{acc}$.  In nearby star-forming regions (\emph{e.g.}, Orion, Taurus, Scorpius), KRAKENS will be able to push the measurements of M$_{acc}$ to the very low values expected for the formation of isolated brown dwarfs (see, \emph{e.g.}, \cite{2009ApJ...696.1589H}).   In more distant star-forming environments, the high sensitivity of KRAKENS on Keck and the smaller angular extent of these regions enable the systematic study of low-mass star formation on a wide scale. These observations can be combined with existing panchromatic surveys from Spitzer and WISE to study the relationship between the accretion lifetime of the systems (assumed to be equal to the gas disk lifetime at r $\le$ 10 AU) and the dust disk dissipation timescale.  These results directly inform our understanding of the initial conditions for the formation and evolution of exoplanetary systems.  

\subsection{Blind Surveys for High-z Galaxies}
\noindent\emph{By George Becker}

%\begin{figure}%[t]
%   \begin{center}
%   \includegraphics[width=\columnwidth]{abs_mag_vs_z.eps}
%   \caption{\footnotesize Top: Limiting absolute magnitude at rest-frame wavelength 1500~\AA\ as a function of redshift.  The redshift range  corresponds to the KRAKENS wavelength range.  Results for seeing limited (1.0'') and tip-tilt corrected (0.4'') observations are shown.  For comparison, the dotted line shows the characteristic absolute magnitude in the UV luminosity function over this redshift range.}
%   \label{fig:galabs}
%   \end{center}
%\vspace{-.2in}
%\end{figure}

%\begin{figure}%[b]
%   \begin{center}
%   \includegraphics[width=\columnwidth]{galaxy_detections.eps}
%   \caption{\footnotesize Cumulative number of galaxies detected in a 5 hour exposure.   Detections are 7$\sigma$ confidence in a single resolution element at $\lambda_{\rm obs} = (1+z)1500$~\AA.}
%   \label{fig:detgal}
%   \end{center}
%\vspace{-.2in}
%\end{figure}

\begin{figure}%[t]
   \begin{center}
   \includegraphics[width=\columnwidth]{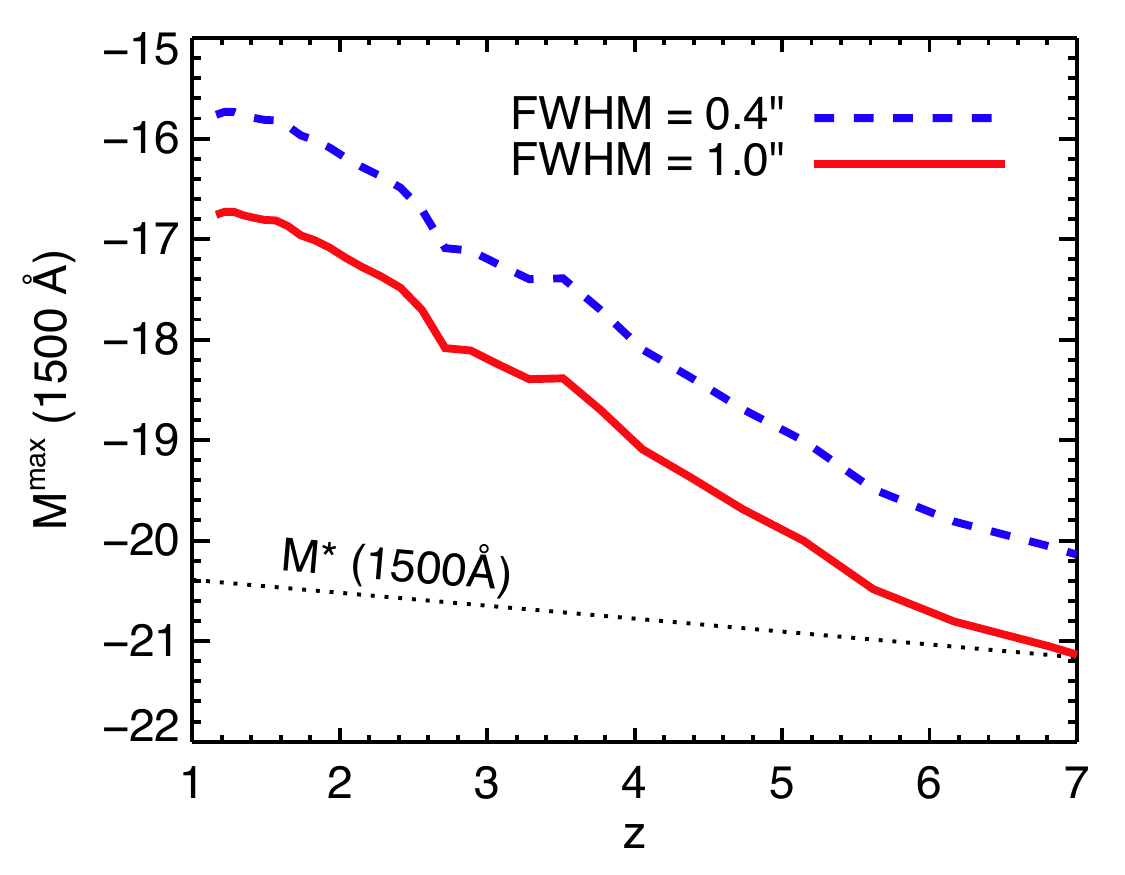}
   \includegraphics[width=\columnwidth]{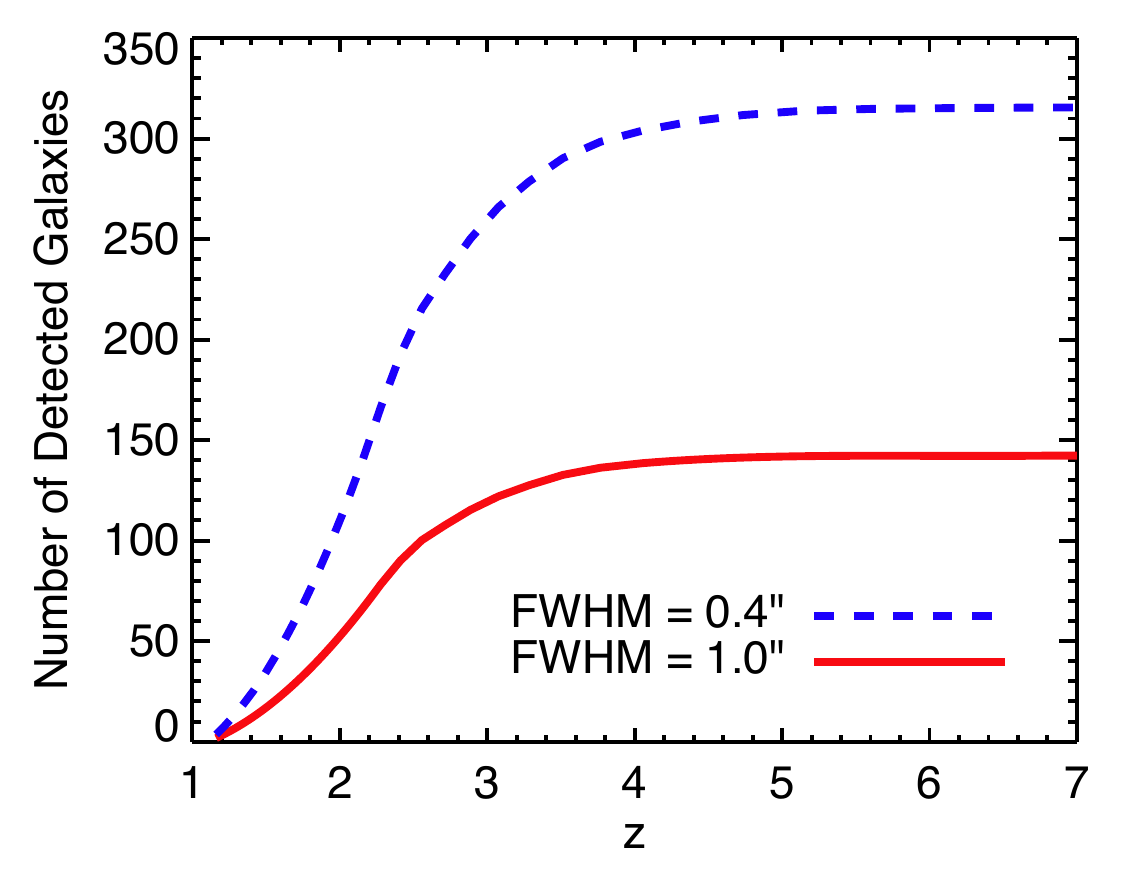}
   \caption{\footnotesize Top: Limiting absolute magnitude at rest-frame wavelength 1500~\AA\ as a function of redshift.  The redshift range  corresponds to the KRAKENS wavelength range.  Results for seeing limited (1.0'') and tip-tilt corrected (0.4'') observations are shown.  For comparison, the dotted line shows the characteristic absolute magnitude in the UV luminosity function over this redshift range.  Bottom: Cumulative number of galaxies detected in a 5 hour exposure.   Detections are 7$\sigma$ confidence in a single resolution element at $\lambda_{\rm obs} = (1+z)1500$~\AA.}
   \label{fig:galabs}
   \end{center}
\vspace{-.2in}
\end{figure}

Its unique combination of wavelength coverage, sensitivity, and field of view will make KRAKENS a powerful tool for extragalactic studies.  Thanks to its broad wavelength coverage, KRAKENS will probe large cosmological volumes in a single pointing.  With a modestly bright ($m < 18$) reference star in the field, moreover, the high time resolution of KRAKENS will enable tip-tilt corrections during processing that should easily give 0.4'' spatial resolution (Nyquist sampled with 0.2'' pixels).  This will deliver a significant gain in sensitivity over seeing-limited large-format IFUs such as KCWI.  In a five-hour exposure with 1.0'' (0.4'') image quality, KRAKENS will deliver low-resolution spectra for $\sim$150 ($\sim$300) galaxies at $z > 1$, as shown in Figure~\ref{fig:galabs}.  Galaxies below the knee of the luminosity function will be detected out to $z \sim 7$, down to 0.02$L^{*}$ at $z=2$.  KRAKENS is therefore well suited for surveys of faint galaxies, including those behind lensing clusters, as well as for targeted observations of brighter galaxies out to the reionization epoch.  It will also efficiently pre-select galaxies over a wide redshift range for slit-based spectroscopic follow-up, as needed for studying the galaxy-IGM interface with quasar absorption lines.

%Technical notes: Sensitivity calculations assume $R_{4000}=30$ and an instrument+telescope throughput of 0.3.   Numbers are based on atmospheric emission and transmission models for Paranal.  

\subsection{Supernovae}
\noindent\emph{By D. Andrew Howell}

\begin{figure*}[ht]
   \begin{center}
   \includegraphics[width=1.5\columnwidth]{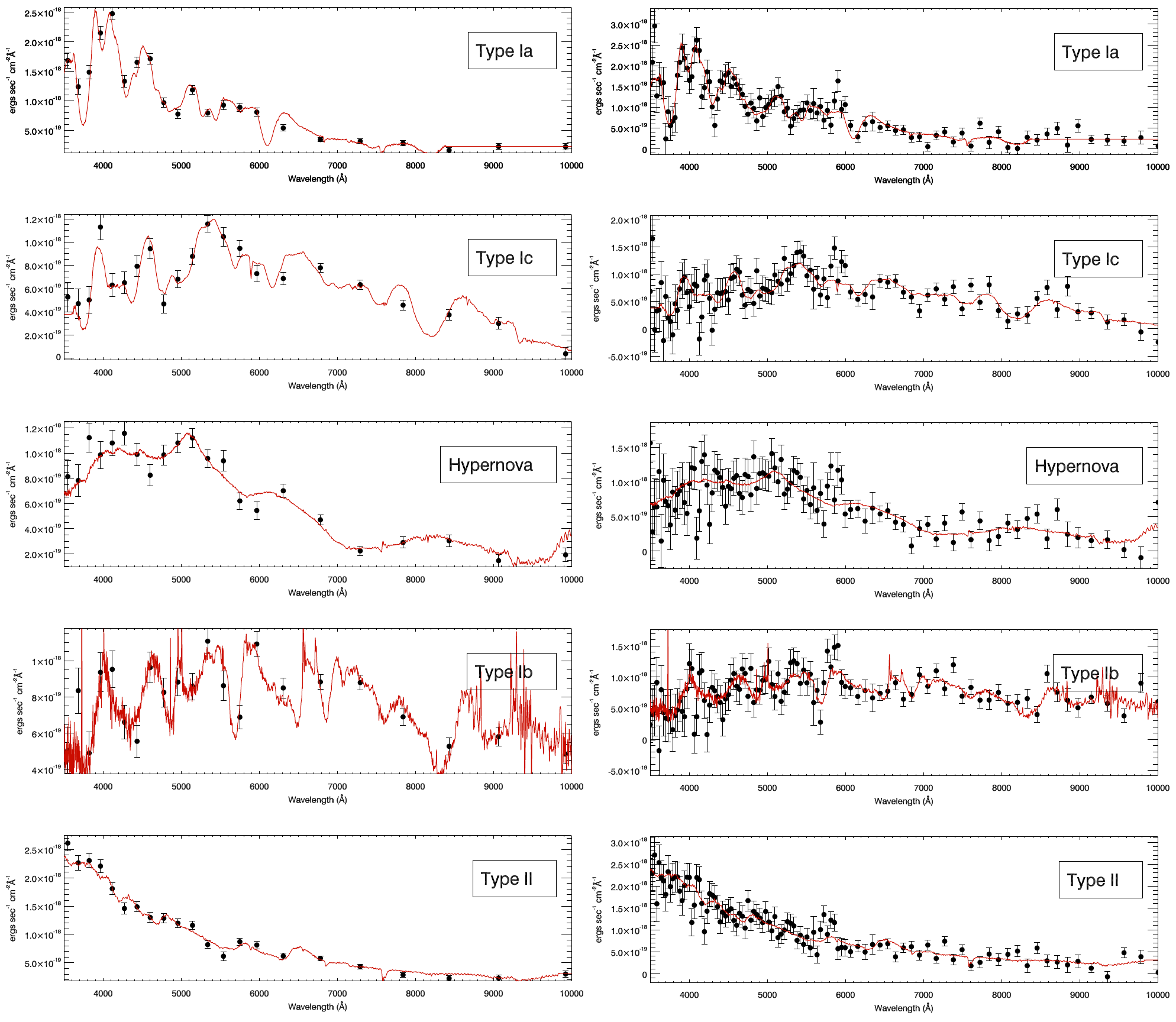}
   \caption{\footnotesize Simulated KRAKENS observations of different types of SNe normalized to m$_V$=24.  The left panel shows observations in the standard R=20 mode for 300s, and the right panel shows simulations in the R=100 mode for 1000s with the dispersive R=100 fore module.  Note that these are only single-epoch spectra.  Since the spectra of each type evolve differently over time, any ambiguities present in a single-epoch snapshot can be resolved by a subsequent observation days later.}
   \label{fig:SN}
   \end{center}
\vspace{-.3in}
\end{figure*}

In a 30 square degree search, the Dark Energy Survey (DES) has discovered more than 1000 Type Ia supernova candidates in its first year of operation, and yet has spectroscopically classified only 70~\citep{Papadopoulos:2015vy}.  With the advent of LSST, hundreds to thousands of SNe per night will be discovered, making the spectroscopic classification problem intractable. 

This problem exists, because, at a typical discovery magnitude of R=24, a spectrum of sufficient signal-to-noise ratio to classify a supernova would take two hours on current-generation Keck instruments (\emph{e.g.} on DEIMOS using the R=600 grating S/N=1 per pixel is achieved in a one hour exposure, but many pixels can be binned).  Using KRAKENS with the R=20 setting, a broadband S/N=25 can be achieved in 300s. 

Figure~\ref{fig:SN} shows the maximum-light spectra (red) of SNe of types Ia, Ic, Ic-broad-line (i.e. hypernova), Ib, and II.  The left panels show the KRAKENS sampling at R=20, and the right panels show the sampling at R=100.  Even with R=20, SN types can be distinguished, sufficient for selecting a sample of SNe Ia for cosmology, or for selecting SNe for subsequent higher-resolution investigations.   At 300s per exposure, using KRAKENS, approximately 100 SNe can be typed in a single night. In reality, there will be a distribution of discovery magnitudes, and supernovae brighten by several magnitudes as they rise to peak, making this a conservative estimate. To achieve the same S/N at R=100 will take approximately 3 times longer, but given the broad nature of SN features (2,000-30,000 km/s), this is sufficient resolution to achieve most scientific objectives.  One can imagine typing 100 SNe during the first night and studying the most interesting 30 in detail on the second night, or even more if the high QE goal is reached.

Only a few hundred SNe have been observed spectroscopically at $z > 0.5$ (see, \emph{e.g.}~\citet{2009A&A...507...85B}), and only a handful of those have sufficient S/N to be used in scientific studies of their features~\citep{2011MNRAS.410.1262W}.  A few night Keck run with KRAKENS would eclipse all previous studies done over several decades.  A sample of 600 well-observed SNe at $z<0.1$ is being currently being constructed from the LCOGT supernova key project.  To search for signs of evolution in the SN population from the early universe to the present day, it is imperative that we construct a similarly-sized SN sample at high redshift.  For example, despite early claims, it does not appear that pair-instability SNe have been discovered in the nearby universe (\emph{e.g.} \citet{2013Natur.502..346N}).  However, they should exist at high redshift.  The mean properties of SNe Ia are also known to be different at low and high redshift~\citep{2007ApJ...667L..37H}.  Metallicity may be to blame and may also be the source of the failure of SNe Ia to correct to the same absolute magnitude in high and low-mass galaxies~\citep{2010MNRAS.406..782S}.  

\noindent\textbf{Shock Breakout to Reveal Progenitors}\\
During a core-collapse supernova, a shock rips through the star, heating it to of order 100,000K, and ultimately breaks through the surface.  This should be visible in the UV to optical bands, and has a duration equivalent to the light-crossing time of the progenitor --- of order an hour for red supergiants, and minutes for stripped-envelope SNe or those from a blue supergiant (\emph{e.g.} SN 1987A).  Immediate cooling then takes place, manifesting as a rapidly falling light curve over the next hours to days, before radioactive decay (or the increasing size of the photosphere) kicks in to power the normal SN lightcurve.  Measuring both the shock breakout and cooling can reveal the size of the progenitor, a great deal about the mass loss history in the hours and days leading to the SN, and even the structure of the star immediately prior to explosion (\emph{e.g.}\citet{Bersten:2012ic}).

While rapidly-falling UV and optical emission has been seen in some SNe, it is always of a longer duration than expected, leading to the conclusion that it is either the after-effects of shock cooling or a shock breakout from an extended wind (e.g. \citet{Ofek:2010gk,2010ApJ...720L..77G,1987ApJ...322L..85D}.  However, proposed large-field-of-view UV surveys (\emph{e.g.} ULTRASAT), or high-cadence optical searches (\emph{e.g.} the HiTS: High cadence Transient Survey) should find bona fide shock breakouts in the 2018 timeframe.  Standard photometric or spectroscopic instruments are ill-equipped to characterize a source that is rapidly changing on the timescale of minutes.  If each exposure is 5 minutes, one may only be able to cycle through the U, B, and V filters once before the shock breakout is over.  And during those exposures, if the light curve is rapidly changing, this information is lost.  However, with KRAKENS a rapidly-changing blackbody can be characterized exquisitely, since each photon is measured in both energy and time.

\subsection{The Unknown}
\noindent\emph{By Jason X. Prochaska}

With the ever-rising stream of optical/IR transients discovered by
all-sky imaging surveys, the potential to uncover new phenomena
increases annually.  From this deluge of sources, researchers will
undoubtedly identify sets of transients whose characteristics (light
curve, color, etc.) differ from all previously known events.  
With the unparalleled sensitivity of the Large Synoptic Survey
Telescope (LSST), it is possible the community will be over-run with a
slew of such `unknown unknowns'.
At $R > 25$\,mag, such
phenomena are likely to overwhelm the capabilities of 
traditional spectrometers on 10-m class telescopes.
These events constitute a ripe discovery space in astronomy where
Keck/KRAKENS could have a major contribution.  

Technically, there are two arenas where KRAKENS on Keck~I would thrive
in the analysis of the unknown unknowns.  These follow from its unique
combination of sensitivity and broad-band spectroscopy together with a
facility capable of time-sensitive observations.
One may expect two usage cases for KRAKENS in this domain:
 (1) rapid response observations to generate a set of SEDs on faint
 sources with steeply declining light curves.  Such observations may
 be critical for the faintest phenomena discovered by LSST.  One may
 even entertain a rapid-fire follow-up campaign where KRAKENS observes
 faint events discovered by LSST in nearly real-time;
 (2) a sequence of deep observations at later times to characterize
 the light curve and search for temporal variations in the SED.  These
 data yield insight into the interactions between the event and its
 surrounding medium and/or variations in the `engine' that has
 generated the event.

\section{Conclusions}

As we have seen in this extensive but not comprehensive science document, KRAKENS will be an extremely powerful instrument across an enormous range of astrophysical topics.  Perhaps more importantly in the current funding climate, it will be a relatively simple and inexpensive instrument with costs in the ${\sim}$2 M range.  With the support of Keck Observatory and the SSC, our team would like to submit a NSF MRI proposal for KRAKENS this fall.

%------------------------------------------------
%\phantomsection
\section*{Acknowledgments} % The \section*{} command stops section numbering
\addcontentsline{toc}{section}{Acknowledgments} % Adds this section to the table of contents

This work would not have been possible without the support of Shri Kulkari, Jason Prochaska, Tom Prince, Greg Hallinan, Sean Adkins, and especially the Keck SSC co-chairs, Judy Cohen and Crystal Martin, as well as generous funding from the W.M. Keck Observatory.

\FloatBarrier

%----------------------------------------------------------------------------------------
%	REFERENCE LIST
%----------------------------------------------------------------------------------------
%\phantomsection

\bibliographystyle{apj}

\renewcommand{\aa} {Astron. \& Ap.}

\newcommand{\aas} {Astron. \& Ap. (Suppl.)}

\newcommand{\aj} {Astron. J.}

\newcommand{\ap} {IEEE Trans. Antennas Propagat.}

\newcommand{\apj } {Ap. J.}

\newcommand{\apjl } {Ap. J. (Lett.)}

\newcommand{\apjs } {Ap. J. (Suppl.)}

\newcommand{\apl } {Appl. Phys. Lett.}

\newcommand{\asc } {IEEE Trans. Applied Superconductivity}

\newcommand{\irmm } {Int. J. IR and MM Waves}

\newcommand{\jap } {J. Appl. Phys.}

\newcommand{\ltp } {J. Low Temp. Phys.}

\renewcommand{\mag } {IEEE Trans. Magnetics}

\newcommand{\mgwl } {IEEE Microwave Guided Wave Lett.}

\newcommand{\mnras } {Mon. Not. Roy. Astron. Soc.}

\newcommand{\mtt } {IEEE Trans. Microwave Theory Tech.}

\newcommand{\pr } {Phys. Rev.}

\newcommand{\prd } {Phys. Rev. D.}

\newcommand{\prl } {Phys. Rev. Lett.}

\newcommand{\rsi } {Rev. of Sci. Instrum.}

\newcommand{\sst } {Supercond. Sci. Tech.}

\newcommand{\icarus } {Icarus}

\newcommand{\nat } {Nature}

\newcommand{\aap } {A\&A}

%----------------------------------------------------------------------------------------

\end{document}